\documentclass[12pt,preprint]{aastex}
\usepackage{emulateapj5,apjfonts}
\usepackage{apjfonts}
\usepackage{graphicx}
\usepackage{amsmath}
\usepackage{color}
\usepackage{onecolfloat}

\newcommand{\lsim}{\lower0.6ex\vbox{\hbox{$ \buildrel{\textstyle <}\over{\sim}\ $}}}
\newcommand{\gsim}{\lower0.6ex\vbox{\hbox{$ \buildrel{\textstyle >}\over{\sim}\ $}}}
\newcommand{\beq}{\begin{equation}}
\newcommand{\eeq}{\end{equation}}

\renewcommand\tabcolsep{0.12cm}

\newcommand{\gammat}{\left<\gamma_t (\theta) \right>}
\newcommand{\DeltaSigma}{\Delta\Sigma (r)}
\begin{document}
\def\head{
\vbox to 0pt{\vss
                    \hbox to 0pt{\hskip 440pt\rm ANL-HEP-PR-XX-XX\hss}
                   \vskip 25pt}

\submitted{The Astrophysical Journal, submitted}

\lefthead{Kwan et al.}
\righthead{Cosmic Emulation}

\shorttitle{Cosmic Emulation}
\shortauthors{Kwan et al.}

\title{Cosmic Emulation: Fast Predictions for the Galaxy Power Spectrum}
\author{Juliana Kwan\altaffilmark{1,2}, Katrin
  Heitmann\altaffilmark{1,3,4}, Salman Habib\altaffilmark{1,3,4},
  Nikhil Padmanabhan\altaffilmark{5}, Earl Lawrence\altaffilmark{6}, 
  Hal Finkel\altaffilmark{7}, Nicholas Frontiere\altaffilmark{1,8} 
  and Adrian Pope\altaffilmark{1,7}}

\affil{$^1$ High Energy Physics Division, Argonne National Laboratory,
  Lemont, IL 60439, USA} 
\affil{$^2$ Department of Physics and Astronomy, The University of Pennsylvania,
  Philadelphia, PA, 19104, USA}
\affil{$^3$ Kavli Institute for Cosmological Physics, The University
  of Chicago, Chicago, IL 60637, USA } 
\affil{$^4$ Mathematics and Computer Science Division, Argonne
  National Laboratory, Lemont, IL 60439, USA} 
\affil{$^5$ Department of Physics, Yale University, 260 Whitney Ave.,
  New Haven, CT 06520, USA}
\affil{$^6$ Statistical Sciences, Los Alamos National Laboratory,
  Los Alamos, NM 87545, USA}
\affil{$^7$Argonne Leadership Computing Facility, Argonne National
    Laboratory, Lemont, IL 60439, USA}     
\affil{$^8$ Department of Physics, The University of Chicago, Chicago,
  IL 60637, USA}

\begin{abstract}
  The halo occupation distribution (HOD) approach has proven to be an
  effective method for modeling galaxy clustering and bias. In this
  approach, galaxies of a given type are probabilistically assigned to
  individual halos in $N$-body simulations.  In this paper, we present a
  fast emulator for predicting the fully nonlinear galaxy-galaxy auto and 
  galaxy-dark matter cross power spectrum and correlation function 
  over a range of freely specifiable HOD modeling parameters. 
  The emulator is constructed using results from 100 HOD models run 
  on a large $\Lambda$CDM $N$-body simulation, with Gaussian Process 
  interpolation applied to a PCA-based representation of the galaxy 
  power spectrum. The total error is currently $\sim 1$\% in the auto
  correlations and $\sim 2$\% in the cross correlations
  from $z=1$ to $z=0$, over the considered parameter range. 
  We use the emulator to investigate the accuracy of various analytic 
  prescriptions for the galaxy power spectrum, parametric dependencies 
  in the HOD model, and the behavior of galaxy bias as a function of 
  HOD parameters. Additionally, we obtain fully nonlinear predictions for
  tangential shear correlations induced by galaxy-galaxy lensing from our
  galaxy-dark matter cross power spectrum emulator. 

  All emulation products are publicly available at http://www.hep.anl.gov/cosmology/CosmicEmu/emu.html.

\end{abstract}
\keywords{Cosmology: $N$-body simulations}}
\twocolumn[\head]

\section{Introduction}

Measurements of galaxy clustering at large scales provide essential
cosmological information, including key inputs to investigations of
dark energy, the growth rate of structure, and neutrino mass. In
particular, observations of two-point clustering statistics, such as
the power spectrum and correlation function of galaxies obtained from
large scale structure surveys, such as the Sloan Digital Sky Survey
(SDSS)/BOSS (Baryon Oscillation Spectroscopic Survey), 
Two-degree Field Galaxy Redshift Survey, and WiggleZ, have been of
particular significance~\citep{pope04, tegmark04, tegmark06, cole05,
  eisenstein05,  parkinson12}. Some of the strongest current
constraints on the nature of dark energy have been derived from
measurements of the Baryon Acoustic Oscillations (BAO)
peak~\citep[e.g.][]{anderson12, anderson14} and redshift space
distortions (RSDs)~\citep[e.g.][]{reid12, reid14}. Aside from the BAO
scale, the amplitude and shape of the galaxy power spectrum and
correlation function provide further cosmological information. In
this case, it is desirable to include as many scales as is
practical in the analysis, however, to do so requires that the
nonlinear regime of structure formation be accurately modeled. As has
been appreciated for quite some time, an essential difficulty is that
galaxies are biased tracers of the underlying density
field~\citep{kaiser84,dekel87}. Because the nature of the bias is
complex and often difficult to unravel, the underlying cosmological
information cannot be straightforwardly extracted.

Modeling the distribution of galaxies remains an enduring problem in
cosmology. $N$-body methods, while extremely successful in capturing the
dark matter distribution at high resolution, do not incorporate the
required baryonic physics for galaxies to emerge out of the large
scale structure self-consistently. Furthermore, the positions of
galaxies do not necessarily follow that of the dark matter, resulting
in a nontrivial bias between statistical measurements of the
clustering patterns between dark matter and galaxies. However, as
mentioned above, accurate modeling of the nonlinear distribution of
galaxies is crucial for extracting cosmological information from large
scale structure surveys and understanding galaxy
formation. Hydrodynamic simulations are still far from attaining the
required degree of maturity needed to provide a complete
first-principles understanding of galaxy formation. For these reasons,
a number of phenomenological approaches -- varying considerably in the
amount of physical input -- have been employed in the continuing quest
to faithfully model galaxy clustering. (For a recent review, see
\citealt{baugh13}.)

The original, and simplest, approach is to assume a (nonlinear,
scale-dependent) fitting function for the bias (defined, say, as the
ratio between the galaxy and the linear or nonlinear matter power
spectrum), combine this with clustering measurements, and marginalize
over the free parameters. Like any such general approach, the problem
is that the fitting form is not necessarily based on a physically
correct model for galaxy formation, and if the form itself is not
sufficiently flexible, this can lead to systematic errors in the
determination of cosmological parameters. (See, e.g., a comparison of
results from different bias models in \citealt{swanson10,
  parkinson12}.)

More detailed models for inferring the location of galaxies can be
obtained by working at the level of dark matter-dominated halos and
subhalos obtained from $N$-body simulations. These methods fall into
three main categories: Halo Occupation Distribution (HOD) modeling,
Subhalo/Halo Abundance Matching (S/HAM) and Semi-Analytic Models
(SAMs). The HOD model is a probabilistic description that aims to
reproduce the statistical distribution of target galaxies on
average. This is achieved by populating dark matter halos with
galaxies as a function of the halo mass. (We discuss HOD modeling more
fully in Section~\ref{sec:HODtheory}.) 

Halo abundance matching is an empirical procedure that involves rank
ordering dark matter halos and subhalos in terms of a particular
characteristic, such as mass or peak circular velocity during its
accretion history~\citep{vale04, conroy06, guo10,  moster10,wetzel10}. 
Similarly, the galaxies are ordered according to an
observable feature, say, luminosity. In this example, the most massive
halo would be matched to the most luminous galaxy, the next most
massive halo assigned the next most luminous galaxy, and so on, until
no galaxies remain. This process ensures that the luminosity function
is exactly reproduced by the synthetic galaxy catalog.

SAMs are the most complex, providing a simplified
accounting of a large number of (baryonic) physical processes,
embedded within $N$-body simulations. They include phenomenological
prescriptions for galaxy formation and associated effects such as gas
cooling, active galactic nuclei and supernova feedback, and star formation, based on the
subhalo and halo formation history, e.g.~\citep{white91, kauffmann93,
  cole94, somerville99, benson03, baugh06, benson10}.

All of these more detailed methods can make predictions for galaxy
clustering (and hence for bias), by using $N$-body simulations and some
number of observational inputs to fix modeling parameters. The results
for the galaxy power spectrum or correlation functions depend on the
modeling parameters, as well as on cosmology. In many cases, it is not
obvious exactly how the final answer depends on the interaction of
these parameters, and an exhaustive sampling of parameter space by
brute force can become computationally very expensive. 

The general problem of efficiently sampling cosmological parameter
space and building fast (essentially instantaneous), accuracy
controlled, simulation-based predictors (``emulators'') for summary
statistics has been addressed via the introduction of the Cosmic
Calibration Framework (CCF). The CCF is based on efficient parameter
sampling strategies coupled to Gaussian Process (GP) based
interpolation and a Markov chain Monte Carlo (MCMC)
sampler~\citep{HHNH, HHHNW}.  The efficiency of the CCF for
reproducing highly nonlinear observables is demonstrated in the
Coyote~\citep{coyote1, coyote2, coyote3} and extended
Coyote~\citep{coyoteplus} emulators for the matter power spectrum,
accurate to 1\% up to $k = 1$~Mpc$^{-1}$ and 3-5\% up to
$k=8.6$~Mpc$^{-1}$, and an emulator for the halo concentration-mass
($c-M$) relation~\citep{kwan13}, accurate to 3\% at $z=0$.

This paper is concerned with providing a means for efficiently
predicting galaxy 2-point statistics within the HOD model, using GP-based
emulation. 
Our method presents a considerable advantage over algorithms that directly sample 
the dark matter halo catalogs~\citep{neistein12}, because these
involve a substantial computational overhead in terms of time and
memory consumption. Moreover, the large scale information in the emulator 
is a product of several realizations of $N$-body simulations to reduce 
finite volume effects; this is not  possible with a single catalogue as 
discussed in~\cite{neistein12}. It is especially powerful because it can be
run on a single processor and each run takes less than a second.

We adopt the HOD model as a first test case for emulation
of galaxy based statistics because it is simple, yet flexible, and
because it is the least demanding in terms of $N$-body simulation
requirements. Using results from a high resolution simulation, we have
populated the halos with galaxies from a sampling design with a 100
different HOD models, measuring the galaxy-galaxy and galaxy-dark
matter power spectra from each model. 
This process is applied to six snapshots between $0 \le z \le 1$ and 
we perform a linear interpolation to obtain additional power spectra at
intermediate redshifts. The emulator is driven by a GP to return either a
galaxy auto or cross power spectrum for arbitrary HOD models within the design.
Sampling from the GP is a fast and accurate means of obtaining a
nonlinear galaxy power spectrum without having to populate a halo
catalog with a new HOD model each time. With the additional input of
source and lens catalogs, the emulator can return the tangential shear
$\gammat$ or the excess surface density, $\DeltaSigma$.

In the following, we discuss our HOD approach, including the parameter
choices, in Section~\ref{sec:HODtheory}. We describe the simulation
underlying this work in Section~\ref{sec:sims} and provide some
details on extracting the galaxy power spectra for the different HOD
models from the simulation. Section~\ref{sec:GP} describes the
emulator construction itself and the tests used for verifying its
accuracy. Section~\ref{sec:analytic} compares the performance of
the emulator to a number of analytic halo models for the galaxy auto and cross
power spectra. An initial set of scientific results based on the new
emulator are reported in Sections~\ref{sec:sens} and \ref{sec:bias},
analyzing the dependence of the galaxy power spectrum on different HOD
parameters and determining galaxy bias for different HOD models. 
In Section~\ref{sec:xi}, we generalize the emulator to configuration 
space. In Section~\ref{sec:gglensing}, we extend the galaxy-dark 
matter cross power spectrum emulator to calculate $\gammat$ and $\DeltaSigma$.
We conclude with a short discussion in Section~\ref{sec:conclusion}.

\begin{figure}[t] %  figure placement: here, top, bottom, or page
   \centering
   \includegraphics[width=\linewidth]{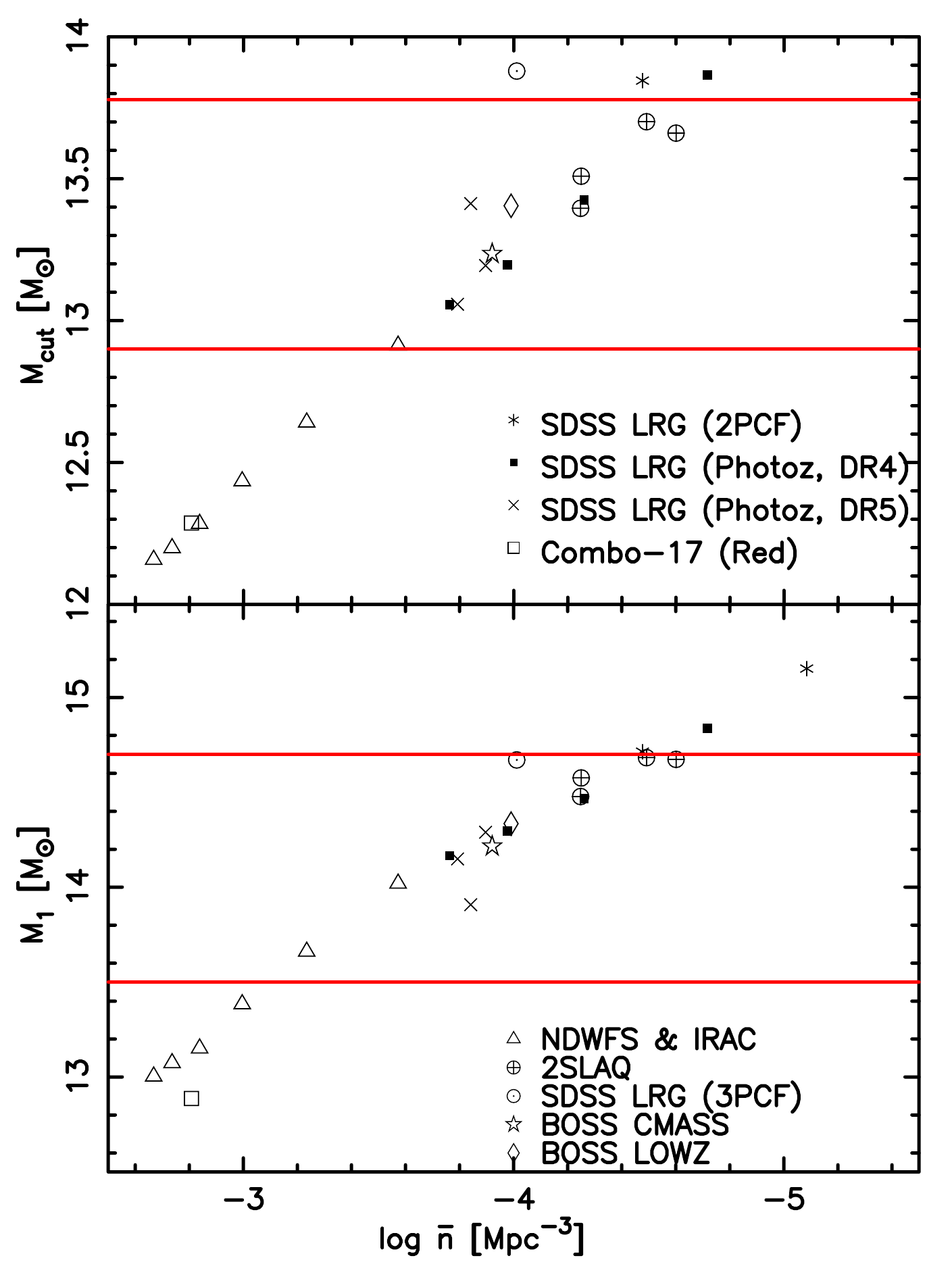} 
   \caption{Range of the HOD emulator for $M_{cut}$ (top panel) and
     $M_1$ (bottom panel) compared to previous HOD analyses.  The
     emulator covers the parameter space between the red lines.  The
     survey data points are taken from Table A1 of~\cite{parejko13}
     and include: ~\cite{zheng09}~[SDSS LRG (2PCF)],
     ~\cite{blake08}~[SDSS LRG (Photo-z, DR4)],
     ~\cite{padmanabhan09}~[SDSS LRG (Photo-z,
     DR5)],~\cite{phleps06}~[Combo-17 (Red)], ~\cite{brown08}~[NDWFS
     \& IRAC],~\cite{wake08}~[2SLAQ],~\cite{kulkarni07}~[SDSS LRG
     (3PCF)], ~\cite{white11}~[BOSS CMASS] and ~\cite{parejko13}~[BOSS
     LOWZ].}
   \label{fig:massranges}
\end{figure}
\section{The Halo Occupation Model}
\label{sec:HODtheory}

The HOD model~\citep{kauffmann97, jing98,
  benson00, peacock00, seljak00, berlind02} has evolved over
time~(cf. \citealt{zheng05}) into a straightforward method for
associating galaxies with halos. The idea behind the HOD model is that
every galaxy is required to be contained within a dark matter halo and
galaxy populations are split into ``centrals", the bright main galaxy
inside the halo, located at the halo center, and surrounding dimmer
``satellite" galaxies. HOD models are calibrated against clustering
observations of sets of target galaxies, allowing for an
interpretation of the measurement in terms of a galaxy population
model for halos. In this sense, the models are not predictive, and, in
principle, have to be tuned to the galaxy population (defined, e.g.,
by color and luminosity) under consideration. (It is also not obvious
that the simple assumption of the halo mass as the master variable is
sufficiently accurate, due to halo assembly bias, as discussed
in~\citealt{gao05}.)

Despite the above caveats, the HOD approach has proven to be very
successful when applied to large scale structure surveys, mostly to
interpret their galaxy populations. These studies have informed us
about the typical host halo mass and the ratio of satellite to central
galaxies for a number of galaxy populations. The HOD model has been
applied to both photometric and spectroscopic galaxy surveys, and
hence galaxy types. These include Luminous Red Galaxies (LRGs), in the
SDSS~\citep{zehavi04, zehavi05, kulkarni07, blake08, zheng09,
  padmanabhan09} and combined 2dF-SDSS LRG and QSO survey
(2SLAQ)~\citep{wake08}, red galaxies from the NOAO Deep Wide Field
Survey (NDWFS), {\it Spitzer} IRAC Shallow Survey~\citep{brown08}
and Combo-17~\citep{phleps06} as well as CMASS~\citep{white11} and
LOWZ~\citep{parejko13} populations from BOSS.

\begin{figure}[htbp] 
   \centering
   \includegraphics[width=\linewidth]{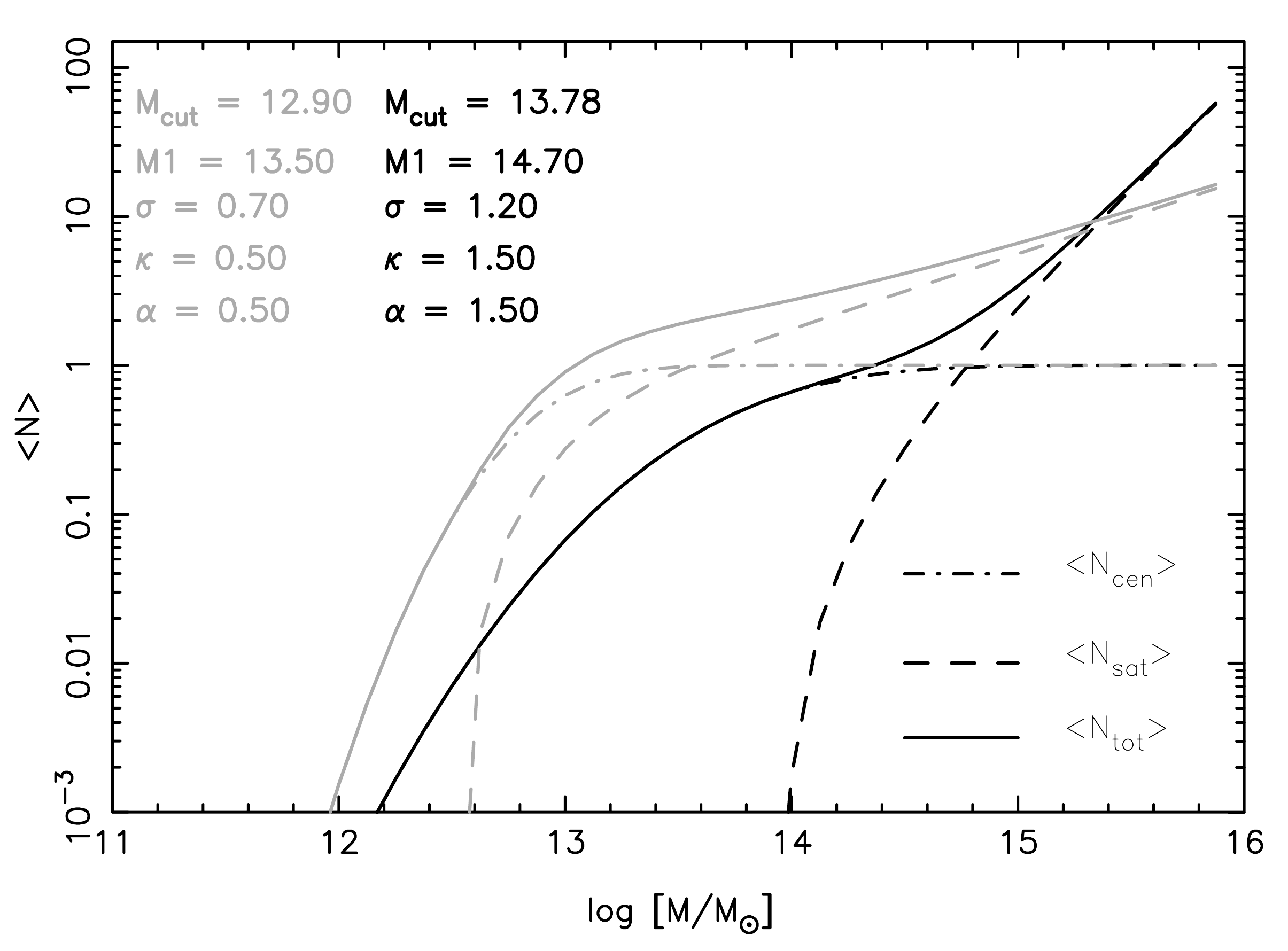} 
   \caption{Mean number of central (dot-dashed) and satellite (dashed)
     galaxies per halo for two extreme HOD models at the edges of the prior
     range of the emulator. The average total number of galaxies per
     halo is shown as a solid curve. The two models shown have the
     lowest (grey) and highest (black) values in the HOD parameter
     space specified in Table~\ref{tab:design}. }
   \label{fig:hod_example}
\end{figure}

In order to be specific, we adopt the HOD model of~\cite{zheng09}
for SDSS LRGs, although other models could easily have been considered. In
this particular case, the average number of central and satellite
galaxies, in a halo of mass $M$, is determined by the following
equations:
\begin{eqnarray}
\left<n_{\rm cen}\right> = \frac{1}{2}\;{\rm{erfc}}\left[\frac{\ln \left(M_{\rm cut}/M\right)}{\sqrt{ 2}
    \sigma}\right],  \\ 
\left<n_{\rm sat}\right> = \left(\frac{M -  \kappa M_{\rm cut}}{M_1}\right)^\alpha.
\label{eqn:zehavi}
\end{eqnarray}
According to the HOD model, each halo must be assigned a probability
of hosting a central galaxy based on the mass of the halo, with the
sharpness of the cutoff mass determined by the parameter $\sigma$.  If
the halo is sufficiently massive to satisfy an additional cut in halo
mass, imposed by $\kappa M_{\rm cut}$, then more galaxies are placed
around the halo center as satellite galaxies; how many of these are
inserted into the halo is controlled by the parameter $\alpha$.  
We assume that the number of satellite
galaxies follows a Poisson distribution with mean
$\left<n_{\rm sat}\right>$. The effect of these parameters on the mean
number of galaxies assigned to each halo are illustrated in
Figure~\ref{fig:hod_example} for two example HOD models.
When calculating the contribution from satellite galaxies, instead of 
drawing dark matter particles from the halo at random based on the
likelihood of hosting a galaxy, if a halo has been determined to host 
a satellite galaxy, each halo particle is assigned a weight according 
to the number of galaxies predicted by the HOD model.  We do
not weight halo particles unless the halo center also has a non-zero 
weight. This is done to reduce the level of shot noise in the power spectra. 
In this scheme, the halo center is given a weight of $\left<n_{\rm cen}\right>$, 
and each particle belonging to the halo has a weight of 
$\left<n_{\rm sat}\right>/N$, where $N$ is the total number of halo particles.

The ranges of HOD parameters that we cover are given in
Table~\ref{tab:design} and illustrated in Figure~\ref{fig:massranges}
with respect to observational values obtained from large scale
structure surveys. Figure~\ref{fig:hod_example} shows the values of
$\left<n_{\rm cen}\right>$ and $\left<n_{\rm sat}\right>$ for the two HOD
models at the extreme ends of the prior range.
The emulator comfortably covers the HOD models used for the analysis
of the recent CMASS BOSS results~\citep{white11} as well as many SDSS
LRG samples at certain redshifts and luminosity cuts.

\begin{table}[ht]
\caption{Prior Range of HOD Model Parameters}
 \label{tab:design}
   \centering
   \begin{tabular}{@{} lcr @{}} 
      12.9      & $ \le \log_{10}(M_{cut} [M_\odot]) \le $& 13.78 \\ 
      13.5  & $\le \log_{10}(M_1 [M_\odot]) \le $ &  14.7  \\
      0.5     & $\le \sigma \le $& 1.2\\
      0.5      & $\le \kappa \le $& 1.5 \\
      0.5 & $\le \alpha \le $&  1.5 \\
   \end{tabular}
\end{table}

\newpage 

The parameter ranges of our emulator are motivated by the galaxy samples 
that we wish to study but ultimately limited by the mass 
resolution of our simulation; we only consider halos with a minimum 
mass cut set by a lower limit of 40 particles per halo, this in turn imposes 
a lower limit on $M_{\rm cut}$ and $\sigma$.  While the smallest halos that 
we populate are actually less massive than the value of $M_{\rm cut}$ 
in that HOD model because $\sigma$ can substantially increase the value of
$\left<n_{\rm cen}\right>$ for low mass halos, we have ensured that these
limits are within the mass resolution of our simulation (discussed
below) by setting an appropriately conservative lower limit on
$M_{\rm cut}$. The upper limit is set mainly by statistical limitations
due to the finite number of high mass halos in the simulation -- a
lower mass cut reduces the amount of noise in the power spectrum, and
is in accordance with current and future galaxy surveys. Galaxy
samples with excessively high $M_{\rm cut}$ and $M_1$ will have a low
number density (high mass halos are rare) and as such there are few
surveys that will target such galaxies. We have checked that the
limits imposed in Table~\ref{tab:design} will miss, at most, 1.6\% of 
galaxies residing in halos below the mass resolution of the simulation.
This translates to an error of $\sim$1\% in the galaxy power spectrum
as calculated from the halo model in the worse case scenario.

\section{$N$-body Simulations}
\label{sec:sims}

Our HOD catalogs are based on an $N$-body simulation with a box-size of
$L = 2100$~Mpc, 3200$^3$ simulation particles and a cosmology similar
to WMAP7: $\Omega_m = 0.2648$ (including both cold dark matter and
baryonic matter), $\Omega_b = 0.0448, n_s = 0.963, \sigma_8 = 0.8$,
and $h = 0.71$. This leads to a particle mass,
$m_p=1.05\cdot10^{10}$M$_\odot$.  The force resolution was set to
$\sim$9~kpc.  Initial conditions were set with the Zel'dovich
approximation, at $z_{\rm in} = 200$.  The simulation was performed using
the HACC (Hardware/Hybrid Accelerated Cosmology Code)
framework~(\citealt{habib09,pope10,habib12}) on the Mira supercomputer
at the Argonne Leadership Computing Facility.

\begin{figure}[htbp]
   \centering
   \includegraphics[width=\linewidth]{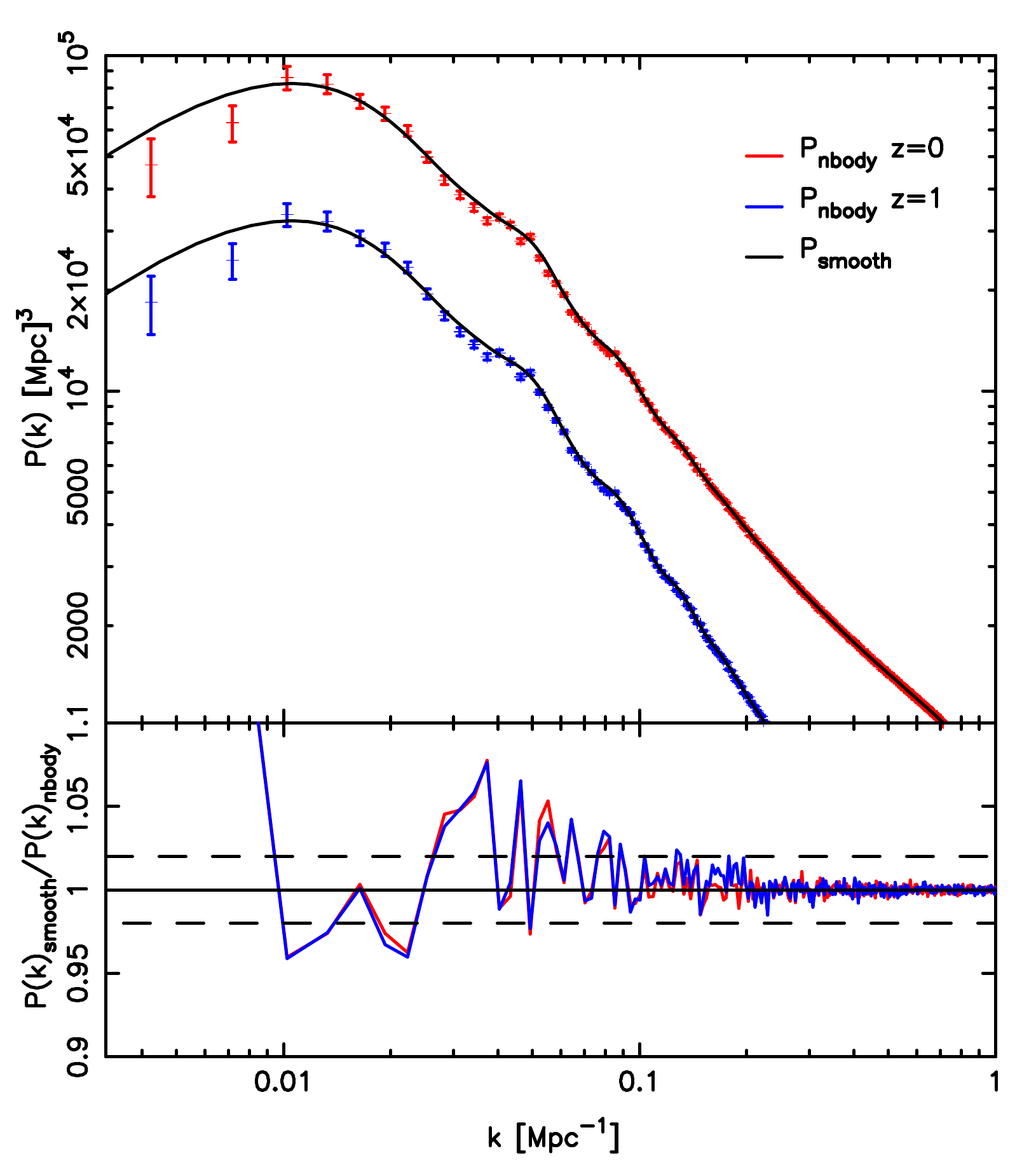} 
   \caption{Matter power spectra measured from the $N$-body simulation
     at $z=0$ (red) and at $z=1$ (blue) with Poisson errors calculated from the 
     number of counts in each bin in $k$. Smoothed matter power spectra
     obtained from an additional 16 PM runs and RPT (as described in
     Section~\ref{sec:smooth_pk}) have also been shown for
     comparison. The lower panel shows the data presented as a ratio;
     the upper and lower horizontal lines indicate a 2\% deviation.}
   \label{fig:pkmcheck}
\end{figure}

To demonstrate the accuracy of our $N$-body simulation, we have shown
the matter power spectrum in comparison to a smoothed average matter
power spectrum in Figure~\ref{fig:pkmcheck}, which was produced by
averaging an additional 15 particle-mesh (PM) simulations combined
with a theoretical matter power spectrum calculated from Resummed
Perturbation Theory (RPT;~\cite{crocce06}), and then smoothed using
a process convolution,  according to the procedure outlined in~\cite{coyote3}.  
Our RPT power spectra were calculated using the perturbation theory package,
Copter~\citep{carlson09}.  The combination of the low resolution
simulations reduces scatter from finite volume effects on the power
spectrum on large scales.  Note that the BAO feature has been enhanced
relative to the simulation as a result of averaging over the
additional realizations. We will later reuse these smoothed matter 
power spectra to obtain smoothed estimates of the galaxy power 
spectra. To avoid finite sampling errors on the very largest scales, 
we model the ratio of the matter power spectrum with respect to the 
galaxy-galaxy and galaxy-dark matter power spectra, 
rather than modeling each separately.

Halos were identified with a Friends-of-Friends (FOF)
algorithm~\citep{einasto84,davis85}. This algorithm groups all
particles that are joined to at least one other particle by a certain
link length, $b$, as belonging to the same halo; approximately
equivalent to requiring a minimum isodensity contour before an
overdensity is considered a halo.  Halo centers are assigned by
identifying the gravitational potential minimum. We chose to use $b = 0.168$, 
since this in rough correspondence with a spherical overdensity (SOD) mass 
of $M_{200}$, reduces halo over-linking, and is also consistent with
other HOD analyses carried out on recent measurements, e.g.
by~\cite{white11} and \cite{parejko13}. The last feature allows for an
easy comparison of results. The smallest halos we consider have at
least 40 particles, leading to a halo mass of $\sim 4.2\cdot
10^{10}$M$_\odot$. At $z=0$, we have a total of $\sim3.4 \cdot 10^{7}$
halos in the simulation and there are $\sim$ 2000 halos with masses in
excess of $9.55\cdot 10^{14}$M$_\odot$, ensuring good statistics for
massive halos.
%NOTE: check the M_200 number above

\subsection{Measuring the Galaxy Auto and Cross Power Spectra from $N$-body
  Simulations}\label{sec:2pt} 

After having identified the halos in the simulations, the next step is
the generation of galaxy catalogs following our HOD prescription
outlined in Section~\ref{sec:HODtheory}.  Varying the set of five HOD
parameters introduced in Table~\ref{tab:design}, we generate 100
different models, arranged in a space-filling Symmetric Latin
Hypercube design as explained in more detail in
Section~\ref{sec:design}.  For each of the 100 HOD models, the halo
catalog is populated with galaxies from which we then measure a galaxy
power spectrum.  The power spectrum is defined as:
\begin{equation}
P(k) = \left<|\delta(k)|^2\right>, 
\label{eqn:pk_def}
\end{equation}
where $\delta = (\rho-\bar{\rho})/\bar{\rho}$ and because we are
interested in characterizing the clustering of galaxies, $\rho$ is the
density of the galaxy field in the Universe. We use a Cloud in Cell
(CIC) deposition on to a 10240$^3$ grid to generate the density field,
followed by a standard power spectrum estimation step using the Fast
Fourier Transform (FFT). We then subtract the Poisson shot noise from 
each galaxy auto power spectrum; under our weighting scheme for the halo
particles, this is defined as $\sum^N_{i=1} w_i/\sum^N_{i=1} w^2_i$, where $w$ is 
weight on each particle as determined by the HOD model. 
For the cross power spectrum, no weighting or shot noise subtraction 
is necessary, since the high resolution of the $N$-body simulation ensures
that the shot noise contribution to the power spectrum is kept small. 

\section{Emulating the Galaxy Auto and Cross Power Spectra}
\label{sec:GP}

Building a prediction scheme, or emulator, for the galaxy power
spectrum, proceeds in three steps: (i) the design step, where we
decide the HOD parameter settings at which to generate the power
spectra, (ii) a smoothing step, where we take the resulting power
spectra and filter discreteness noise caused by the finite number of
galaxies in our catalogs, (iii) the interpolation step, where we build
a GP model to generate predictions at new points in the
HOD parameter space, leading to the final emulator.  Next, we describe
each of these steps in detail, followed by a rigorous testing
procedure to verify the accuracy of our new emulator.

\subsection{Design Strategy}
\label{sec:design}
The distribution of models in the five-dimensional HOD parameter space
-- the emulator design -- is determined by a Symmetric Latin Hypercube
to cover the maximum amount of parameter space with the fewest
models. The technique for generating such a design is detailed
in~\cite{coyote2} (including many references); the basic premise is
that it is a space-filling design such that in any given
two-dimensional projection of the full five-dimensional space, the
models are approximately evenly sampled. The challenge is to determine
a sufficiently large sample of models such that the target accuracy
can be achieved without wasting computational time by oversampling.

Initially, we tried a set of 100 HOD models that span the range given
in Table~\ref{tab:design}.  The number of models chosen to cover the
parameter space is determined by performing a series of tests in which
we vary the number of design points used from 25, 50, to 100 and build
a toy emulator for each set using halo model predictions for the HOD
power spectrum as a proxy model~\citep[see][for example]{cooray02}
because these can be generated quickly. According to the halo model,
the galaxy power spectrum is given by:
\begin{equation}
P_{gg}(k) =  P^{1h}_{gg}(k)+ P^{2h}_{gg}(k);
\label{eqn:halopk}
\end{equation}
and 
\begin{eqnarray}
P^{1h}_{gg} = \int n(m) \frac{\left< N_{\rm gal} \left(N_{\rm gal} -
      1\right)|m\right>}{\bar{n}} | u(k|m) |^2\; dm
      \label{eqn:2h} \\ 
P^{2h}_{gg} =  P_L \left[\int n(m) \; b(m) \frac{\left< N_{\rm gal} | m
    \right>}{\bar{n}} u(k|m) \;dm\right]^2,  
\label{eqn:1h}
\end{eqnarray}
where $n(m)$ is the halo mass function, $\bar{n}$ is the mean density
of galaxies and $u(k|m)$ is the dark matter mass profile in Fourier
space, $P_L(k)$ is the linear matter power spectrum and $b(m)$ is the
bias.  
Similarly, for the galaxy-dark matter cross power spectrum, we can write:
\begin{equation}
P_{gm}(k) =  P^{1h}_{gm}(k)+ P^{2h}_{gm}(k);
\label{eqn:halopk_cross}
\end{equation}
and 
\begin{eqnarray}
%hmf[3*j+2]*pow(10.,hmf[3*j])/rho*u[i]*(u[i]*nsat[j]+ncen[j])/nbar;
P^{1h}_{gm} = \frac{1}{\bar{n}}\int n(m) \frac{m}{\bar{\rho}}  \left[  |u(k|m)|^2  \left<N_{\rm sat} |m \right > + u(k|m) \left<N_{\rm cen} |m \right >\right]  dm 
      \label{eqn:2h_cross} \\ 
P^{2h}_{gm} =  P_L \left[\int n(m) \; b(m) \frac{\left< N_{\rm gal} | m
    \right>}{\bar{n}} u(k|m) \;dm\right],  
\label{eqn:1h_cross}
\end{eqnarray}
where $\bar{\rho}$ is the mean density of dark matter in the Universe. 

\begin{figure}[htbp] 
   \centering
   \includegraphics[width=\linewidth]{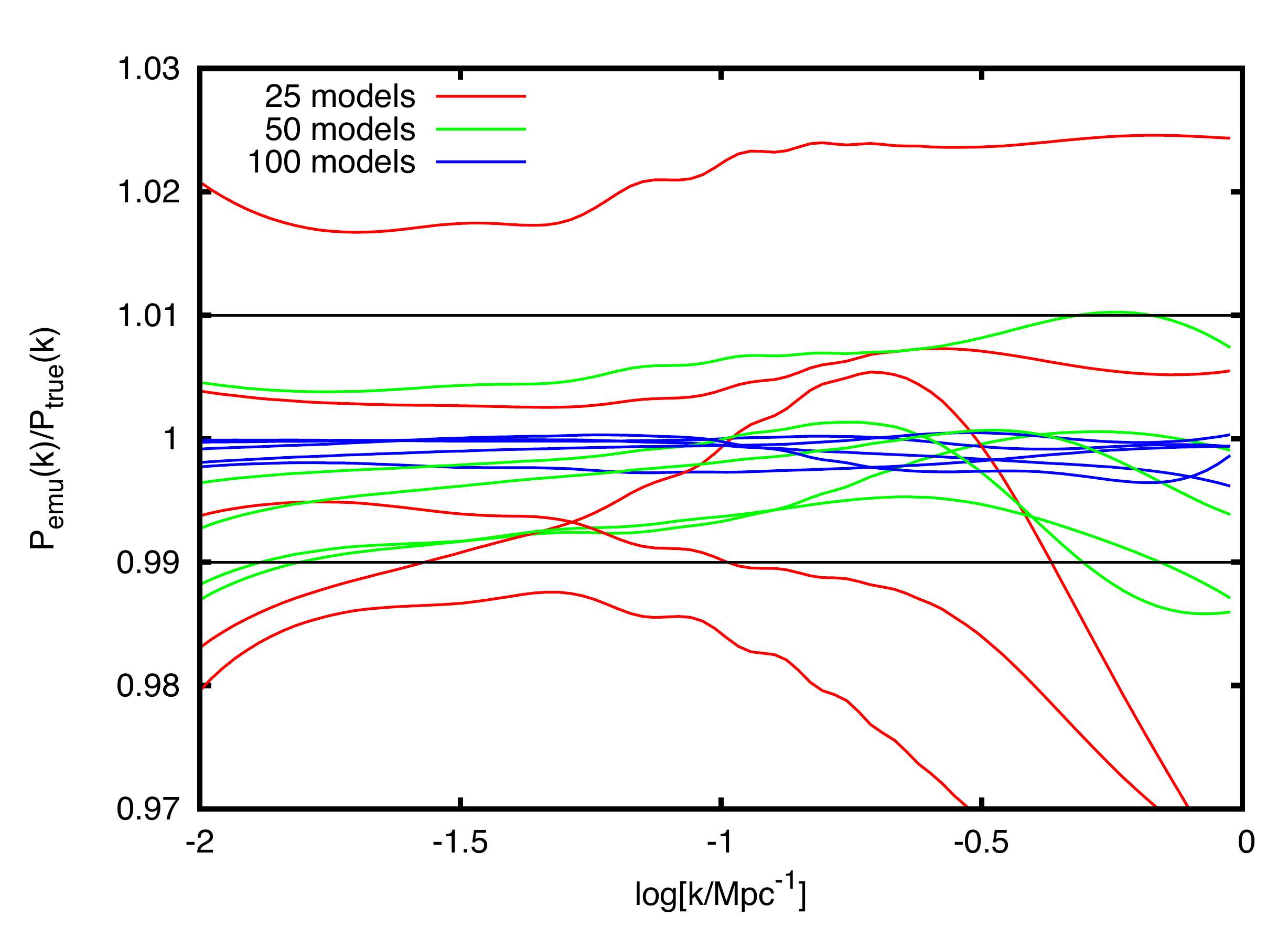} 
   \caption{Accuracy test on the toy emulators built from halo model
     proxies. The aim is to estimate the number of models needed to
     cover the space of five HOD parameters with percent level
     accuracy. We built three emulators based on linear theory HOD
     models with 25 (red), 50 (green) and 100 (blue) design points and
     use these to predict the power spectrum of five models not
     included in the designs, which we denote $P_{\rm true}(k)$.  The
     horizontal black lines denote our targeted accuracy of 1\%.}
   \label{fig:conv}
\end{figure}

Since there exist analytic prescriptions or fitting formulae
for many of these terms, we can calculate these quantities much more
readily compared to using galaxy catalogs from $N$-body simulations. The
accuracy checks on these toy emulators are shown in
Figure~\ref{fig:conv}, in which we have selected five models not
included in any of the designs and compared the predictions from each
emulator to these. We found the proxy model easily achieves subpercent
level accuracy with only 100 models. However, the response surface can be more
complicated in the fully nonlinear case than in the simplified proxy
model. In fact, we required 100 HOD models at redshifts, $z = 0 -
0.66$, but 149 models for redshift $z=1$ to assure percent level
accuracy in the final product. Unfortunately, the proxy model cannot
fully account for the effect of shot noise in the galaxy power spectra
and for the range of HOD models we considered, the shot noise was
sufficiently different across the parameter space to require a closer
sampling at high redshift where the halos are sparser. The estimate
of the resultant accuracy of our emulator is verified by our later
{\em a posteriori} tests (Section~\ref{sec:holdouts}) carried out on
the full emulator.

\subsection{Smoothing the power spectrum}\label{sec:smooth_pk}

In this section, we discuss the smoothing process used to convert the
power spectra into noise-free estimates suitable for emulation. The
galaxy auto and cross power spectra generated from the simulation contain measurement
noise, because of finite volume effects and discrete sampling of
Fourier modes. For the GP to function properly, we do not want to
model noisy estimates of quantities as this would interfere with the
ability of the GP to smoothly vary across the parameter space because
of the random noise included with each function. Therefore we smooth
the power spectra before they are used to condition the GP.  We
require that this effective filtering introduces errors of no more
than $\sim$ 1\% percent into the power spectrum measurement.

\begin{figure}[htbp]
   \centering
   \includegraphics[scale=0.55] {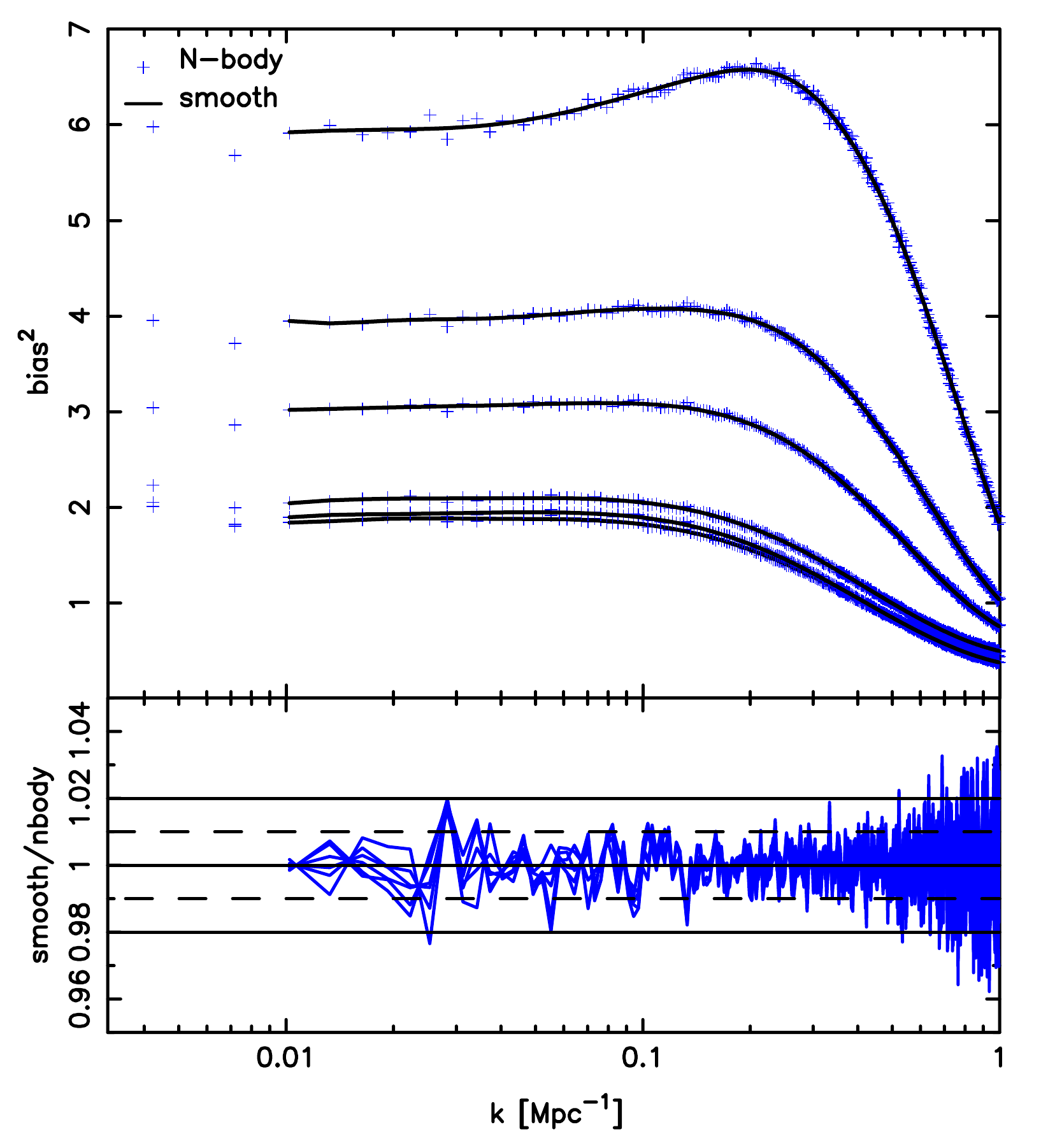}
   \caption{Example HOD model results obtained after applying the smoothing 
     process at six redshifts. The top panel shows the ratio $P_{gg}/P_{m}$
     measured from both the $N$-body simulation (blue crosses) and the smoothed HOD 
     models (black curve).  The bottom panel shows the ratio between the two, with 
     dashed and solid lines indicating 1\% and 2\% error bands respectively.}
   \label{fig:check_pk}
\end{figure}

Our process for smoothing for the power spectrum proceeds 
in the following steps: 

\begin{enumerate}
\item We measure the set of matter, galaxy-dark matter and galaxy-galaxy power spectra from 
the $N$-body simulation using the same sized FFT grids.
\item We take the ratio between the matter and the galaxy-galaxy and galaxy-dark matter
power spectra to give the bias. This removes much of the scatter from
finite volume effects, seen in Figure~\ref{fig:pkmcheck} on large
scales, in the power spectra. 
\item We then perform a basis spline on the binned power spectra. 
  The bias is a simple enough function such that we can use a
  basis spline of order 4 with 10 coefficients evenly spaced throughout
  the $k$-range to capture the dependence of the bias on scale. 
  The variance in the bias is sufficiently low such that the 
  spline is able to capture the shape without much error as demonstrated
  in Figure~\ref{fig:check_pk}. 
\item The power spectra are then recovered by multiplying the bias with 
 a smoothed estimate of the dark matter power spectrum. This is obtained from 
 the same procedure that was used in Figure~\ref{fig:check_pk}. 
 In~\cite{coyote3}, it was shown that the smoothing process, which uses
 additional information from the linear regime in the form of 15 low resolution
 simulations and perturbation theory, correctly captures the matter power 
 spectrum to 1\% accuracy. 

\end{enumerate}

In Figure~\ref{fig:check_pk}, we show an example HOD galaxy-galaxy power spectrum
with parameters, $M_{\rm cut}$ = 13.7086, $M_1$ = 13.4515, $\sigma$ =
0.6061, $\kappa$ = 0.9444 and $\alpha$ = 1.1364, chosen at random,
after applying all the steps in the smoothing process. The results 
shown in the figure demonstrate that visually, there are no discernible 
defects in the galaxy power spectrum caused by our smoothing procedure
and that the basis spline is sufficiently complex to fit the data points.

\subsection{Gaussian Process Modeling}
Once the smoothed power spectra have been obtained at the
design points, a Gaussian Process model is conditioned on these
results, and can be interrogated to provide power spectrum predictions
for any set of parameters chosen to lie within the prior range of
Table~\ref{tab:design}.

The GP is a family of non-parametric, Gaussian distributed functions
about a set of input points. The GP returns a function whose behavior
is obliged to satisfy the input points at high accuracy. Our Gaussian
model exists in parameter space, and not in $k-$space; i.e. the GP
does not model each $k$ bin individually but rather the function
as a whole over the entire parameter set.  Overfitting is avoided by
supplying a covariance function that regulates the complexity or
``smoothness" of the function returned by the GP. This is achieved by
controlling the relationship between each model in parameter space in
terms of a distance metric. It is important that the underlying
response surface mapped by the GP varies smoothly with the parameters
-- this requires the absence of sudden discontinuities as we move from
one model to another with similar parameters. In most cosmological
applications, this is not an issue, as most two-point statistics are
quite well behaved when the underlying parameters are
changed. However, we often do not know in advance the exact
dependencies and degeneracies that exist in parameter space,
particularly if the problem is nonlinear. For this reason, the form of
the covariance function is parameterized with a set of
hyperparameters.  These are determined by maximizing the likelihood
of these parameters given the simulation data, which we carry 
out via an MCMC process.

Our procedure for setting up the GP closely follows the method
outlined in ~\cite{HHNH},~\cite{HHHNW} and~\cite{coyote2}. We have
only briefly summarized the process here, because we are not so much
concerned with the use of GPs for precision cosmology, but the
application of GPs to the particular problem at hand. We refer the
interested reader to the earlier papers for further details.

Once the GP is fully specified, we can draw a function, constrained to pass
through the design points, at any point in the parameter space that
satisfies the covariance function.  This process is no more computationally
expensive than calculating of the inverse of the covariance matrix 
with the new model included.

\subsection{Testing the Emulator}\label{sec:holdouts}
\begin{figure}[htbp]
%   \centering
   \hspace{-1cm}
   \includegraphics[scale=0.39] {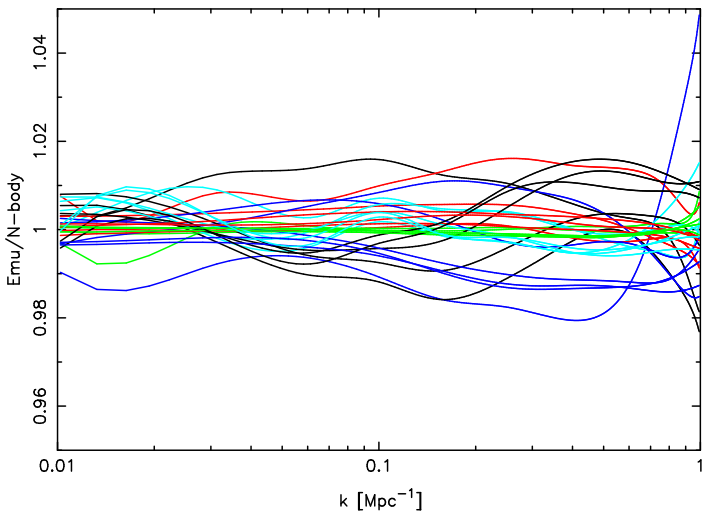}
   \caption{Accuracy test for $P_{gg}(k)$ : five HOD power spectra at each redshift
     are predicted by the emulator and compared to the same models directly
     measured from the $N$-body simulation not included in the original
     design. Each model is represented by a different color. The HOD power spectra
     returned by the emulator are within $\sim$ 1\% of the smoothed
     $N$-body results on the scales $0.01 \leq k \leq 1$~Mpc$^{-1}$ for the
     models tested. }
   \label{fig:holdout_gg}
\end{figure}

\begin{figure}[htbp]
  % \centering
   \hspace{-1cm}
   \includegraphics[scale=0.4] {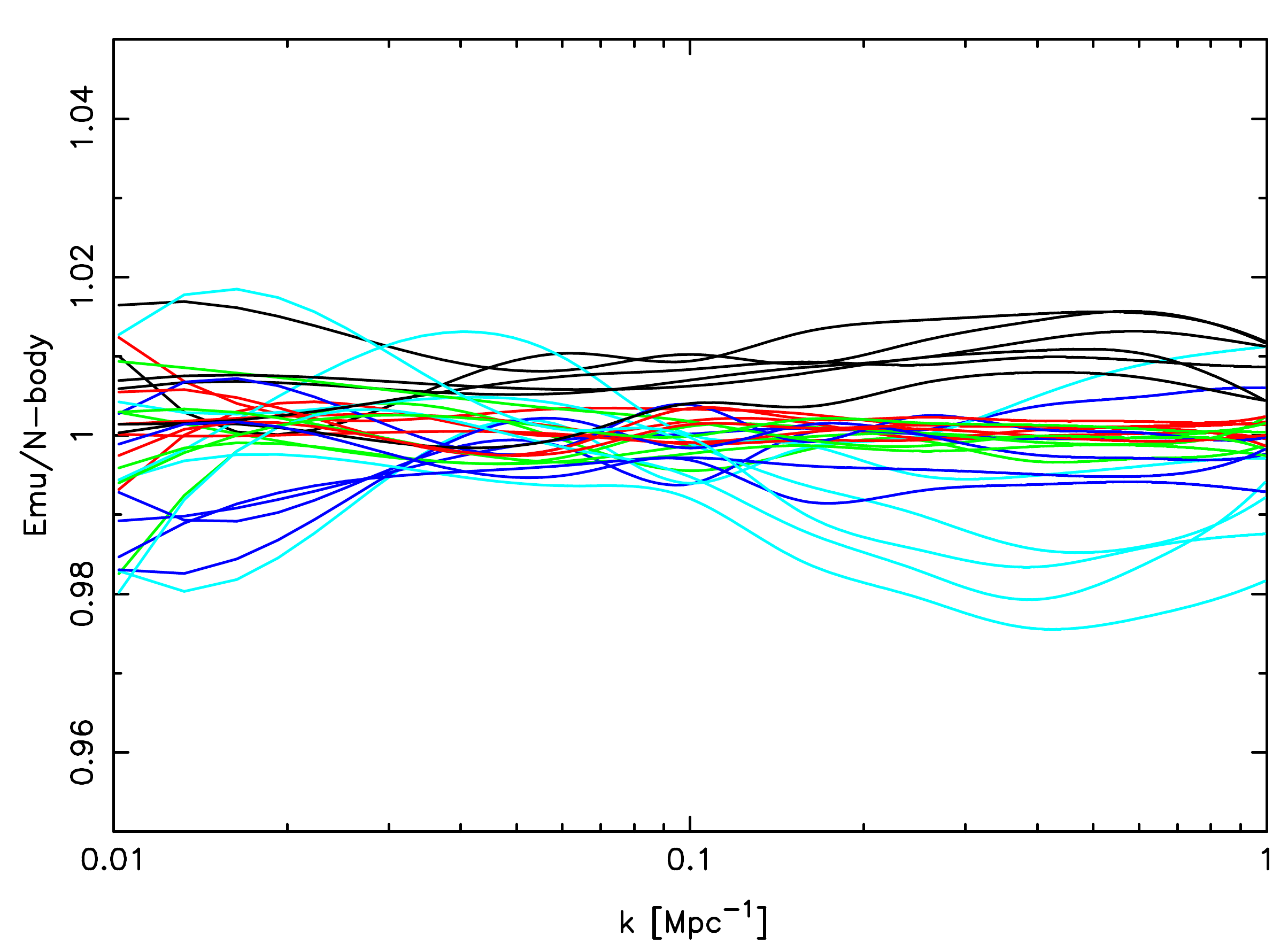}
   \caption{Accuracy test for $P_{gm}(k)$ at all redshifts. As in
     Figure~\ref{fig:holdout_gg}, we test the 
     emulator against five new HOD models drawn from the $N$-body 
     simulation that were not included in the original design.
     Each color represents the same set of HOD parameters drawn
     from the within the parameter range at each of the six redshifts
     used to build the emulator.}
   \label{fig:holdout_gm}
\end{figure}

In this section, we test the accuracy of the emulator by comparing the
power spectrum generated by the emulator to HOD models directly
sampled from our N-body simulation but not included in the conditioning
of the GP. We apply the same smoothing process, described in
Section~\ref{sec:2pt}, to these new HOD models.  We repeat this test
on both $P_{gg}(k)$ and $P_{gm}$ at each of the six redshift slices used
to construct the emulator.  In Figures~\ref{fig:holdout_gg}, and~\ref{fig:holdout_gm}, 
we show the results of these tests, demonstrating that the emulators are indeed accurate
to $\sim 1$\% and $\sim 2$\% respectively, over the range $0.01 \leq k \leq 1$~Mpc$^{-1}$.
These accuracy limits are well below the accuracy requirement on $P_{gm}$  
to extract HOD constraints from galaxy-galaxy lensing data for current experiments.
Our test models are chosen at random to span the full range of parameters. 
Generally, the emulator should perform better near the center of the
design and worse at the edges of the Latin hypercube, simply because
there are a limited number of models that support the design edge.
This is seen in some of the blue curves in Figure~\ref{fig:holdout_gg}, 
particularly at $z=1$, which is poorly reproduced from $k \sim 0.8$ Mpc$^{-1}$ 
onwards because it lies on a corner of the design space and because 
galaxies from this HOD utilize the most massive halos i.e. $M_{\rm cut} = 13.78$ and 
$M_1 = 14.7$ and hence require the most shot-noise 
subtraction.

Note that the errors in Figure~\ref{fig:holdout_gg} 
are percent level in the fully nonlinear case rather than below sub-percent 
level as in Figure~\ref{fig:conv} because the response surface is more 
complicated with nonlinear structure formation and there are additional 
contributions to the error budget in smoothing and shot noise.

%\begin{figure}[htbp]
%   \centering
%   \hspace{-1cm}
%   \includegraphics[scale=0.4] {figures/holdout_z1.pdf}
%   \caption{Accuracy test for $P_{gal}(k)$ at $z=1$. As in
%     Figure~\ref{fig:holdout_z1}, we test the emulator against five
%     new HOD models drawn from the N-body simulation that were not
%     included in the original design.}
%   \label{fig:holdout_z1}
%\end{figure}

\section{Comparison to Analytic Models}\label{sec:analytic}
We now compare the accuracy of our emulator to analytic predictions of
the HOD power spectrum. These models are based on summing the 2-halo
and 1-halo contributions to the galaxy power spectrum. The relevant
equations for the most basic halo model (see, e.g.,
\citealt{cooray02}) are listed in Section~\ref{sec:design}
(Equations~\ref{eqn:halopk} -- \ref{eqn:1h}). The 2-halo term
(Equation~\ref{eqn:2h}) describes galaxy pairs in two different halos,
while the 1-halo term (Equation~\ref{eqn:1h}) arises from the galaxy
pairs that occupy the same halo.  There have been many revisions to
this model and we consider two of the most popular, the \cite{zheng04}
and~\cite{tinker05} models, which we will call Z04 and T05 respectively. 

There are four ingredients to these models: the halo profile, the 
concentration-mass relation, the halo bias and the halo mass function. 
Whenever possible, we take the most recent fitting functions that are 
the most widely accepted in the literature to model these four quantities.  
We use an NFW profile to describe the distribution of galaxies in a halo. 
For the concentration-mass relation, we use~\cite{bhat13}, which was
calibrated on a $\Lambda$CDM cosmology that closely resembles our
simulation in its $\Omega_m$ and $\sigma_8$ values.

To model the large scale halo bias, we use the following fitting
function from~\cite{tinker10} 
\begin{equation}
b(\nu) = 1 - A\frac{\nu^a}{\nu^a+\delta_c^a} + B\nu^b + C\nu^c
\label{eqn:tinkerbias}
\end{equation}
where $A = 1+0.24y\exp\left[-(4/y)^4\right]$, $a = 0.44y-0.88$,
$B=0.183$, $b=1.5$, $C = 0.019+0.107y+0.19\exp\left[-(4/y)^4\right]$,
$c = 2.4$ and $y=\log_{10}\Delta$ and $\Delta = 200$. For our FOF
catalogs with $b=0.168$, $\Delta = 200$ is the most appropriate
background overdensity value considered in~\cite{tinker10}, who used halo
catalogs identified with a SOD finder. 
Indeed,~\cite{tinker10} report that a good agreement was
found in the measured values of the large scale bias between FOF halo
catalogs with $b=0.168$ and a SOD catalog of $\Delta = 200$,
despite the differences in the methodology and effects of aspherical
FOF halo isodensity contours~\citep{lukic07}. To this end, we also 
use the mass function from~\cite{tinker08}; although this is also calibrated 
on SOD halos, the normalization of the mass function is consistent 
with Equation~\ref{eqn:tinkerbias} for the halo bias such that 
$\int b(\nu) n(\nu) \, d\nu = 1$ and we would like to limit our analysis 
to only include model ingredients that are publicly available.

Our implementation of both the Z04 and T05 models use the same halo
mass function, concentration-mass relation and the same expression 
for the large scale, linear halo bias. There are, however, two points
on which the models differ; firstly the treatment of halo exclusion and 
secondly, the functional form assumed for the evolution of the halo bias
as a function of scale. The Z04 model imposes halo exclusion by setting 
the upper integration limit, $M_{\rm lim}$, on the 2-halo term to avoid counting 
contributions from two overlapping halos. This is done by requiring 
$M_{\rm lim} = 4/3\pi (r/2)^3\rho_c \Omega_m\Delta$, 
where $r$ is the radius of the halo, such that no other halo residing within 
half of the radius of a halo of mass, $M_{\rm lim}$, can be considered. 
T05 extends this halo exclusion model by allowing halos to be ellipsoidal 
and by modelling the distribution of the ratio of their major to minor axes.
We can then calculate the effect of non-spherical halo alignments on the 
2-halo term thusly:
\begin{eqnarray}
& & P^{2h}_{gg}(k,r) = \frac{1}{\bar{n}'^2} P_m(k)\int^\infty_0 n(M_1)\left<N_{\rm gal}|{M_1}\right> b(M_1,r) u(k|M_1) \; dM_1 \nonumber\\
& & \qquad \qquad \int^\infty_0 n(M_2) \left<N_{\rm gal}|{M_2}\right> b(M_2,r) u(k|M_2) p(y) \; dM_2
\label{eqn:tinker2h}
\end{eqnarray}
where $\bar{n}'$ is the reduced number density and 
$p(y) = 3y^2-2y^3, y = (x-0.8)/0.29$ and is the probability of 
non-overlapping halos as calibrated from $N$-body simulations. The 
function $p(y)$ is bounded such that, when $y < 0$, $p(y) = 0$ 
and when $y>1$, $p(y) = 1$. Equation~\ref{eqn:tinker2h} then 
requires a Hankel transform to remove the remaining dependence on 
scale and is reweighted thusly:
\begin{equation}
1+\xi^{2h}(r) = {\left(\frac{\bar{n}'}{\bar{n}}\right)}^2\left[1+\xi'^{2h}(r)\right], 
\label{eqn:tinkerxi}
\end{equation}
where $\xi'^{2h}(r)$ is just the Hankel transform of $P^{2h}_{gg}(k,r)$ from
Equation~\ref{eqn:tinker2h}. 
Because the double integral in Equation~\ref{eqn:tinker2h} is time consuming to evaluate, we use the 
$\bar{n}'$ matched limit as suggested in T05. This involves calculating
the reduced number density as 
\begin{equation}
\bar{n}'^2 = \int^\infty_0  n(M_1) \left<n_{\rm gal}|M_1\right> \;dM_1 \int^\infty_0 n(M_2) \left<n_{\rm gal}|M_2\right> \; dM_2, 
\label{eqn:tinkernbar}
\end{equation}
then finding the value for the upper limit on the integral over halo mass
that gives an equivalent number density to Equation~\ref{eqn:tinkernbar} 
and replacing the $M_{lim}$ in the Z04 model with this value. 

The halo bias in the T05 model is given by: 
\begin{equation}
b^2(M,r) = b^2(M)\frac{\left[1+1.17\,\xi_m(r)\right]^{1.49}}{\left[1+0.69\,\xi_m(r)\right]^{2.09}}
\label{eqn:radialbias}
\end{equation}
where $b^2(M)$ is the expression in Equation~\ref{eqn:tinkerbias} from~\cite{tinker10}
and $\xi_m$ is the matter correlation function. Unfortunately, the form 
of the scale dependence of the halo bias used in Z04 is not explicitly 
written out (it is only stated that it is calibrated to $N$-body simulations) and 
so we can only use the large scale asymptotic bias in the 2-halo term.

\begin{figure}[htbp] 
   \centering
   \includegraphics[width=1.15\linewidth]{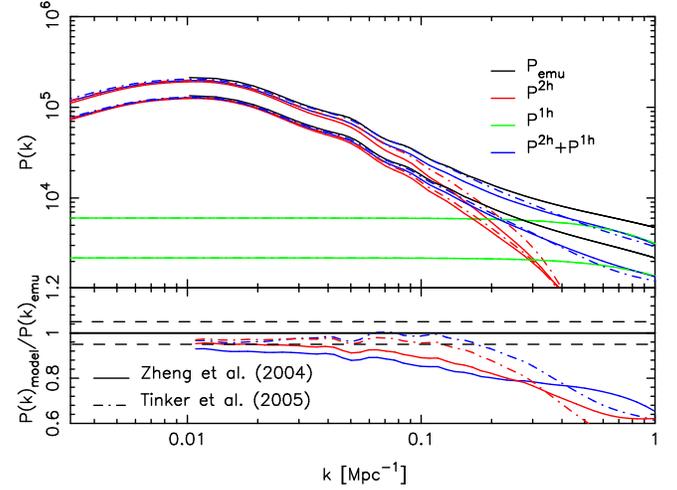} 
   \caption{The auto and cross power spectra from the HOD emulator compared 
     to the \cite{zheng04} (Z04; solid) and~\cite{tinker05} (T05; dotted-dashed) 
     analytic models. We have split the total power spectrum from each of 
     these models into their 2-halo (red) and 1-halo (green) components. 
     The bottom panel shows the error in the analytic models compared to the
     emulator for the galaxy-dark matter cross power spectrum in red and
     the galaxy-galaxy power spectrum in blue. 
     The large scale bias in the T05 model is reproduced quite well at the
     $\sim$ 5\% level up to $k\sim 0.17$ Mpc$^{-1}$, but then there are
     deviations of up to 40\% on smaller scales.}
   \label{fig:halo_model_comparison}
\end{figure}

In Figure~\ref{fig:halo_model_comparison}, we compare the Z04 and T05 
models (solid and dotted-dashed, respectively), calculated using the halo model
components described above, against our HOD emulator. We chose a 
random set of model parameters within our acceptable parameter range.  
On large scales, both power spectra agree to $\sim$10\% 
(better than $\sim5$\% in the T05 model), then the Z04 model starts to deviate at 
$k\sim0.06$ Mpc$^{-1}$, as the scale dependent bias becomes important and 
the contribution from the 1-halo term is inadequate at this scale to substantially
increase the amplitude of the total power spectrum.
In contrast, the T05 model remains accurate to $\sim 5\%$ down to $k\sim0.17$ 
Mpc$^{-1}$. By allowing for non-spherical halos, there are additional contributions to 
the 2-halo term from halos that are fortuitously aligned along their minor axes 
and $P^{2h}(k)$ in the T05 model is boosted relative to the Z04 model. 
This results in the overall power spectrum having a better fit to the simulations. 
Note that the 1-halo contributions are the same for both models. 

The evolution of the halo bias with scale makes a significant contribution in the T05 model
in matching the amplitude of the power spectrum to the HOD emulator by
boosting the linear halo bias. However, there are indications that the modeling of
the scale dependent bias is not ideal. If we neglect the scale dependence 
in the halo bias in Equation~\ref{eqn:tinker2h}, the T05 model is accurate to
10\% down to $k \sim 0.3$ Mpc$^{-1}$ but the discrepancy between this version
of the T05 model and the emulator is approximately constant with $k$. 
This suggests that the scale dependent bias would not be necessary (for this set 
of HOD parameters) if the large scale linear bias was better captured by~\cite{tinker10}. 
This implies that accurately characterizing the halo bias is a worthwhile 
endeavour if we are to improve the halo model.

Past the quasi-linear scale, neither model can be trusted to derive accurate
constraints on cosmology or the HOD parameters as the shape of the 
power spectrum is significantly biased at the 20-40\% level. Unfortunately, 
this sort of halo model approach is only as good as its 
constituent fitting functions and the accuracy of these may be severely 
restrictive and dependant on the cosmology, volume, mass resolution etc.
of the simulations used to calibrate them. Furthermore, the evaluation of 
these models is very slow, (the double integral in Equation~\ref{eqn:tinkernbar} 
is particularly time consuming as is the transformation to configuration space
for Equation~\ref{eqn:tinkerxi}); each model can take up to a minute to compute, 
compared to less than a second for the emulator. 

Nonetheless, these models give an intuitive understanding of the HOD
power spectrum via the halo model and its 2-halo and 1-halo
contributions.  Furthermore, while undoubtedly more accurate, by
construction, the emulator can only operate within a certain parameter
range, while, in principle, the halo model can be less restricted. 
Unfortunately, some of the ingredients of the halo model are also 
calibrated on $N$-body simulations, which can carry their own 
assumptions, such as the choice of a particle cosmology or 
in the implementation of a technique e.g. FOF versus SOD halo finding. 

\begin{figure*}
\centering
\includegraphics[width = 0.5\columnwidth, angle=270]{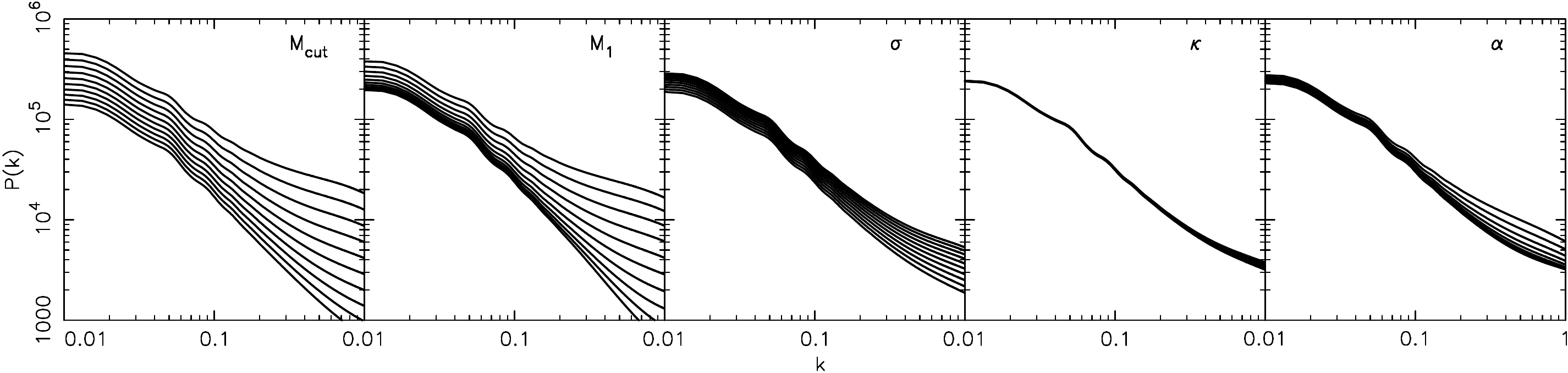}
\caption{Emulated galaxy auto power spectrum at $z=0$. The five HOD
  parameters $M_{\rm cut}, M_1, \sigma, \kappa$ and $\alpha$ are varied.
  A single parameter is changed at a time in each panel, while the
  other four parameters are kept fixed at the midpoint of the
  parameter space. We divide the range of each parameter into ten
  evenly spaced bins, these are the values fed into the HOD
  emulator. }
  \label{fig:sensitivity}
\end{figure*}

\newpage
\section{Parameter Sensitivities}
\label{sec:sens}
Now that we are in possession of an HOD emulator, we can smoothly vary
each HOD parameter in turn to investigate parametric degeneracies and
other effects on the galaxy auto and cross power spectra. In
Figure~\ref{fig:sensitivity}, we explore the effect of changing each
parameter on the galaxy-galaxy auto power spectrum. We divide the parameter space
into ten evenly spaced bins for each HOD parameter in turn, while
keeping all the other parameters fixed at the center of the design,
i.e. $M_{\rm cut} = 13.35, M_1 = 13.8, \sigma = 0.85, \kappa = 1$ and $\alpha
= 1$. The resultant series of multiple power spectra are plotted in
Figure~\ref{fig:sensitivity}.

The results shown in Figure~\ref{fig:sensitivity} demonstrate that the
parameters that most strongly affect the HOD power spectrum are
$M_{cut}$ and $\alpha$, while $\kappa$ only minimally affects the
power spectrum over our parameter range.  The parameter, $M_{\rm cut}$,
has the greatest influence in determining which halo will host a
central galaxy. Since the majority of galaxies in our HOD models are
centrals, it follows that $M_{\rm cut}$ has the greatest effect on the HOD
power spectrum, especially on the linear bias. 

\begin{figure}[htbp]

 \hspace{-1cm}
   \includegraphics[width=1.2\linewidth] {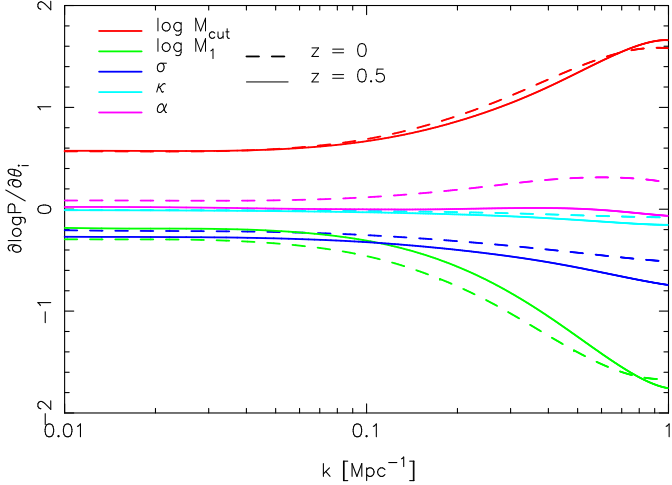}
   \caption{
     Derivatives of the galaxy-galaxy power spectrum with respect to the 5
     HOD parameters, $\theta_i = \{M_{\rm cut}$ (red), $M_1$ (green),
     $\sigma$ (blue), $\kappa$ (cyan) and $\alpha$ (magenta) $\}$.  
     We compute each partial derivative at the midpoint
     of the design at two redshifts, $z=0$ (dashed) and $z=0.5$
     (solid).}
   \label{fig:derivs}
\end{figure}

\begin{figure}[htbp]

 \hspace{-1cm}
   \includegraphics[width=1.2\linewidth] {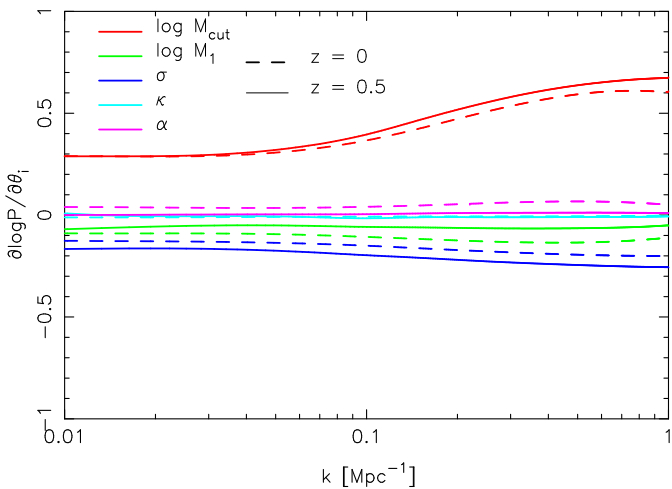}
   \caption{As in Figure~\ref{fig:derivs}, but for the galaxy-dark
   matter cross power spectrum. Note the change in scale on the y-axis.}
%   \caption{Derivatives of the galaxy-dark matter cross power spectrum 
%     with respect to the 5 HOD parameters, $\theta_i = \{M_{cut}$ (solid), $M_1$ (dashed),
%     $\sigma$ (dot dashed), $\kappa$ (dotted) and $\alpha$ (dot-dot
%     dashed) $\}$.  We compute each partial derivative at the midpoint
%     of the design at two redshifts, $z=0$ (grey) and $z=0.5$
%     (black).}
   \label{fig:derivs_cross}
\end{figure}

In Figures~\ref{fig:derivs} and~\ref{fig:derivs_cross}, we have calculated $\partial \log P(k)/\partial
\theta_i$ from the emulator as a function of wave number, (here $\theta_i = \{ M_{\rm cut},
M_1, \sigma, \kappa, \alpha\}$) to demonstrate the degeneracies
between the HOD parameters. 
The Fisher information matrix assess how well a parameter can be 
measured from a particular statistic and is defined as follows: 
\begin{equation}
F_{ij} = - \left < \frac{\partial^2 log f}{\partial\theta_i\partial\theta_j}\right>
\label{eqn:fisher}
\end{equation}
From \cite{tegmark97}, we can approximate the Fisher matrix with 
\begin{equation}
F_{ij} \approx 2\pi \int^{k_{\rm max}}_{k_{\rm min}} \left(\frac{\partial \log P(k)}{\partial\theta_i}  \right)\left(\frac{\partial\log P(k)}{\partial\theta_j} \right) w(k) d\log k
\label{eqn:fisher_approx}
\end{equation}
Assuming the same survey window, $w(k)$, Figures~\ref{fig:derivs} and~\ref{fig:derivs_cross},
give an insight into how well these HOD parameters can be measured 
from the auto and cross power spectra respectively at two redshifts
$z=0$ (grey) and $z=0.5$ (black). 

Figure~\ref{fig:derivs} shows that all the HOD parameters are degenerate on large scales, 
since they all show a similar relationship with $k$ up to $k = 0.05$~Mpc$^{-1}$. This implies that they are all 
capable of shifting the linear, large scale asymptotic bias; but higher $M_{cut}$ and
$\alpha$ values will increase the bias as the galaxy catalog will be
populated from higher mass halos and more satellite galaxies, whereas
increasing $M_1$ will decrease the bias, as a catalog with a higher
$M_1$ will contain fewer satellite galaxies, if all other parameters
are kept fixed. The parameter, $\sigma$, widens the mass cut on the
central galaxies to accept more low mass halos and increasing $\sigma$
reduces the linear bias.  At smaller scales up to $k = 1$~Mpc$^{-1}$,
$M_{\rm cut}$ affects the shape of the HOD power spectrum more strongly
than any other parameter. At these scales, the power spectrum is still
dominated by the two-halo term, so the additional satellite galaxies
produced by having a larger value of $\alpha$ contributes less
clustering than does $M_{\rm cut}$.  As in Figure~\ref{fig:sensitivity},
$\kappa$ does very little to change the shape of the power
spectrum. We have also investigated these relationships at $z=0.5$, as
shown in Figure~\ref{fig:derivs}. By this redshift, the number of
massive halos has been greatly reduced compared to $z=0$. This in turn
reduces the influence of $M_1$ and $\alpha$ on the HOD power spectrum,
which are only active parameters if $M \gtrsim
10^{14}$~$M_{\odot}$. Conversely, $\sigma$ and $M_{\rm cut}$ become more
influential at higher redshift. We have also tried varying the central point 
in parameter space from which we calculate the derivatives of the parameter
values. We found our conclusions to be qualitatively unchanged when the 
`midpoint' is shifted to either edge of the parameter range, although the 
overall amplitudes of $\partial\log{\rm P}/\partial\theta_i$ may be more or 
less pronounced.

For the galaxy-dark matter cross power spectrum, Figure~\ref{fig:derivs_cross}
shows that the dominant HOD parameter that determines its amplitude and shape 
is the mass cut off for the central galaxies, $M_{\rm cut}$. We can also expect to 
constrain, $\alpha$, which controls how many satellite galaxies to insert into
each halo, much more readily than the typical mass of the halo hosting the 
galaxies, $M_1$, while $\sigma$ and $\kappa$ make very little difference 
to the shape and bias of the cross power spectrum.

Our results agree with \cite{parejko13}, who found similar dependencies on the 
shape of the projected HOD correlation function, $w_p$, although their study
probes much smaller scales than our HOD power spectrum emulator.

\section{Nonlinear Bias}
\label{sec:bias}
\begin{figure}[htbp]
 %  \centering
%   \hspace{-1cm}
   \includegraphics[width=\linewidth] {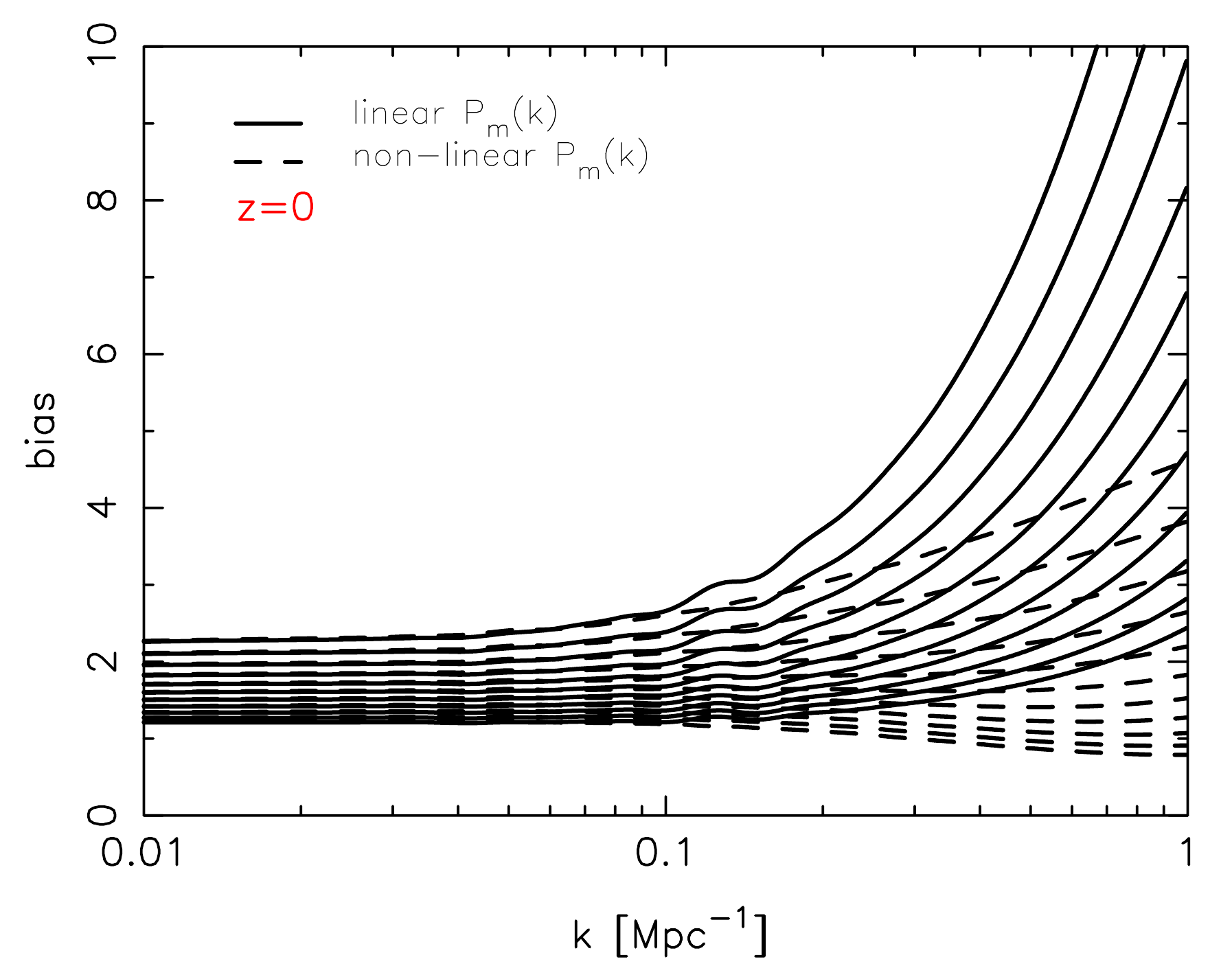}
   \caption{Nonlinear galaxy bias determined from the HOD emulator
     calculated using the linear $P_m(k)$ (solid) and nonlinear
     $P_m(k)$ (dashed) at $z=0$. We have varied $M_{\rm cut}$ between the
     maximum and minimum parameter ranges to produce 10 different
     curves.  The other HOD parameters are kept constant at their
     midpoint values.}
   \label{fig:bias_z0}
\end{figure}

\begin{figure}[htbp]
 %  \centering
%   \hspace{-1cm}
   \includegraphics[width=\linewidth] {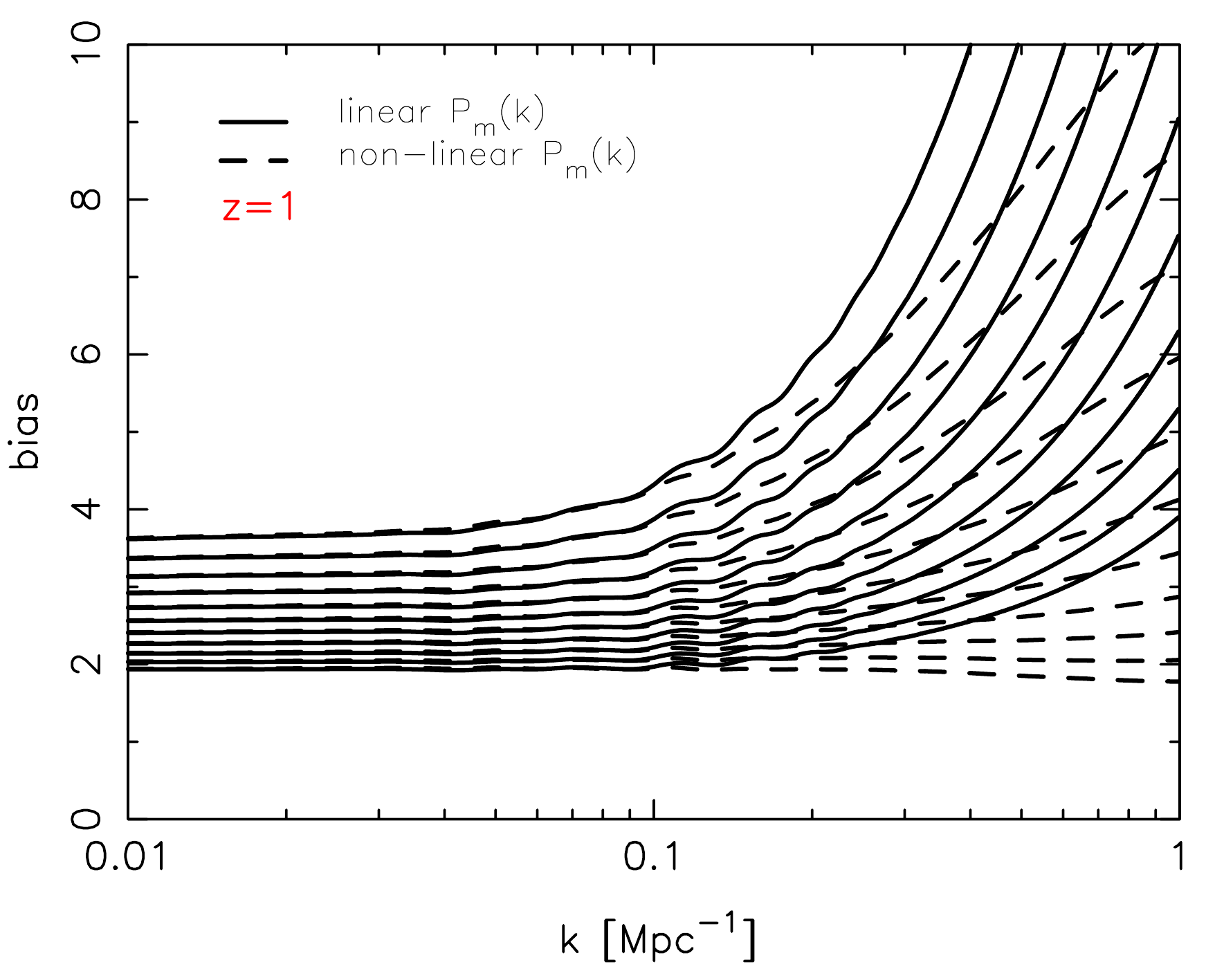}
   \caption{Galaxy bias, following Figure~\ref{fig:bias_z0}, but at $z=1$. }
   \label{fig:bias_z1}
\end{figure}

We now investigate the nonlinear galaxy bias with our HOD emulator.
This is a difficult quantity to model analytically beyond the large
scale, linear limit and our emulator offers a means of easily
accessing nonlinear predictions for the galaxy bias.  We define the
galaxy bias as follows:
\begin{equation}
b(k) = \sqrt{\frac{P_{gg}(k)}{P_m(k)}},
\label{eqn:bias}
\end{equation}
where $P_m(k)$ can be chosen to be either the linear or nonlinear
matter power spectrum, defining two notions of galaxy bias.  In
Figures~\ref{fig:bias_z0} and~\ref{fig:bias_z1}, we show the evolution
of the galaxy bias as a function of scale at $z=0$ and $z=1$,
respectively. As in Figure~\ref{fig:sensitivity}, we have divided the
parameter range into 10 bins, but in this section, we allow only
$M_{cut}$ to change, since this is the parameter that the HOD power
spectrum is most sensitive to, as shown previously in
Section~\ref{sec:sens}.

Figures~\ref{fig:bias_z0} and~\ref{fig:bias_z1} show that the
nonlinearity of the bias increases with redshift when the HOD model is
kept the same. The scale dependence of the bias in relation to the
nonlinear matter power spectrum is quite moderate for the models with
a low $M_{\rm cut}$ and is approximately linear until $k \sim
0.1-0.2$~Mpc$^{-1}$.  The scale dependence is stronger at higher
redshift, but this is because a galaxy catalog at this redshift with
the same HOD parameters will contain rarer halos, i.e. there were much
fewer $10^{15}$~M$_\odot$ halos at $z=1$ than $z=0$ and so these are
more biased with respect to the matter density field.

In evaluating power spectra where the density field is reconstructed
from mass points, there is an unavoidable shot noise contribution due
to the finite mass resolution. At the highest $k$ values considered
here, the shot noise in the matter power spectra is insignificant,
because the particle Nyquist wave number in the simulation is
sufficiently large~\citep[see][for detailed evaluations and
tests]{coyote1}. A similar situation exists for the galaxy field: the
preferential sampling of halos amongst the dark matter distribution
introduces an element of shot noise in the galaxy power spectrum. We
subtract a Poissonian shot noise term, proportional to $1/\bar{n}$ in
keeping with current analyses of observational data,
e.g~\cite{anderson12}.  But halos are biased tracers that tend to
follow the highest peaks of the dark matter density field, and by
placing galaxies inside these, we have inherently chosen positions
that are not a fair sample of the entire density field and so the
generation of shot noise is not entirely a Poisson process in the
galaxy power spectrum.  It is important to note that the galaxy shot
noise will make a substantial contribution to the bias shown in
Figures~\ref{fig:bias_z0} and~\ref{fig:bias_z1} on small scales.


\section{Configuration Space}\label{sec:xi}
We now consider the HOD power spectra in configuration space.  Certain
features, e.g. the baryon oscillations, are more prominent in
configuration space than in Fourier space; additionally, the
correlation function can be more readily measured from galaxy surveys
than can the power spectrum, making the correlation function
a more attractive quantity to model. %check this 

The correlation function is related to the power spectrum via the 
following transformation: 
\begin{equation}
\xi(r) = \frac{1}{2\pi^2} \int k^2 \; P(k)\; j_0(kr) \:dk,
\label{eqn:pktoxi}
\end{equation}
where $j_0$ is the spherical Bessel function. Performing this integral 
is numerically challenging because of the highly oscillating integrand. 
Nonetheless, we attempt this brute force approach to obtain a reference
correlation function which can be compared to more sophisticated methods. 
In order for the integral to converge, we introduce a smoothing term by 
multiplying the integrand with a damping factor of $\exp{(-k^2\sigma^2)}$ where
$\sigma = 0.5$. 
We tried two other methods in addition to the brute force approach, 
a quadrature formula to approximate integrals over Bessel functions 
introduced by~\cite{ogata05} and applied to the correlation
function by~\cite{szapudi05} and FFTlog, an algorithm that performs 
Fast Fourier or Hankel transforms over logarithmically spaced intervals
~\citep{hamilton00}. 

\begin{figure}[htbp] 
   \centering
   \includegraphics[width=1\linewidth]{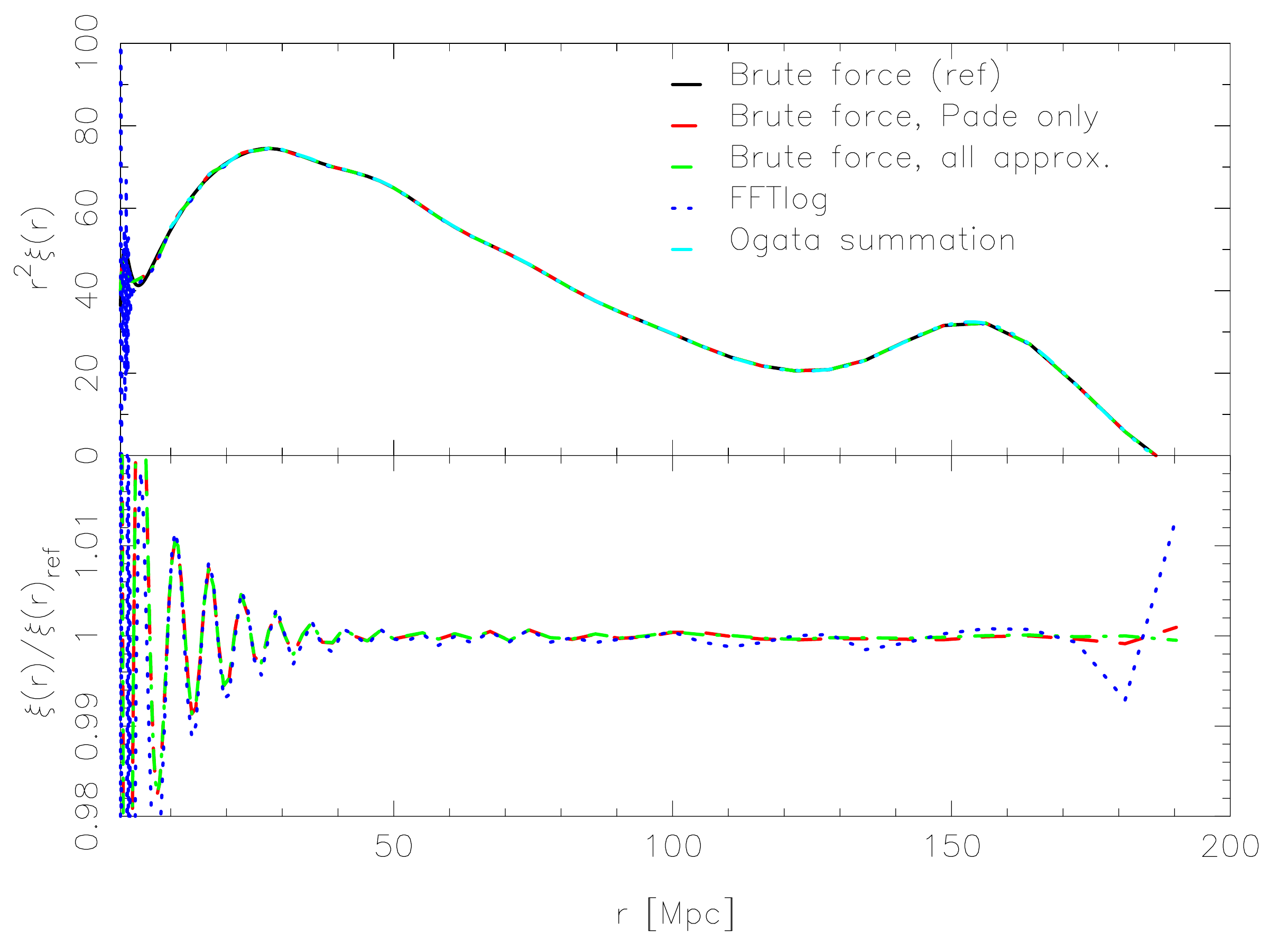} 
   \caption{The matter correlation function produced by different
   schemes for calculating the Hankel transformation from Fourier space.
   This shows that the quadrature formula of~\cite{ogata05}
   and the extrapolations that we employ to calculate the correlation function
   from the power spectrum only introduces less than 1\% error. }
   \label{fig:xi_transform_matter}
\end{figure}

\begin{figure}[htbp] 
   \centering
   \includegraphics[width=1\linewidth]{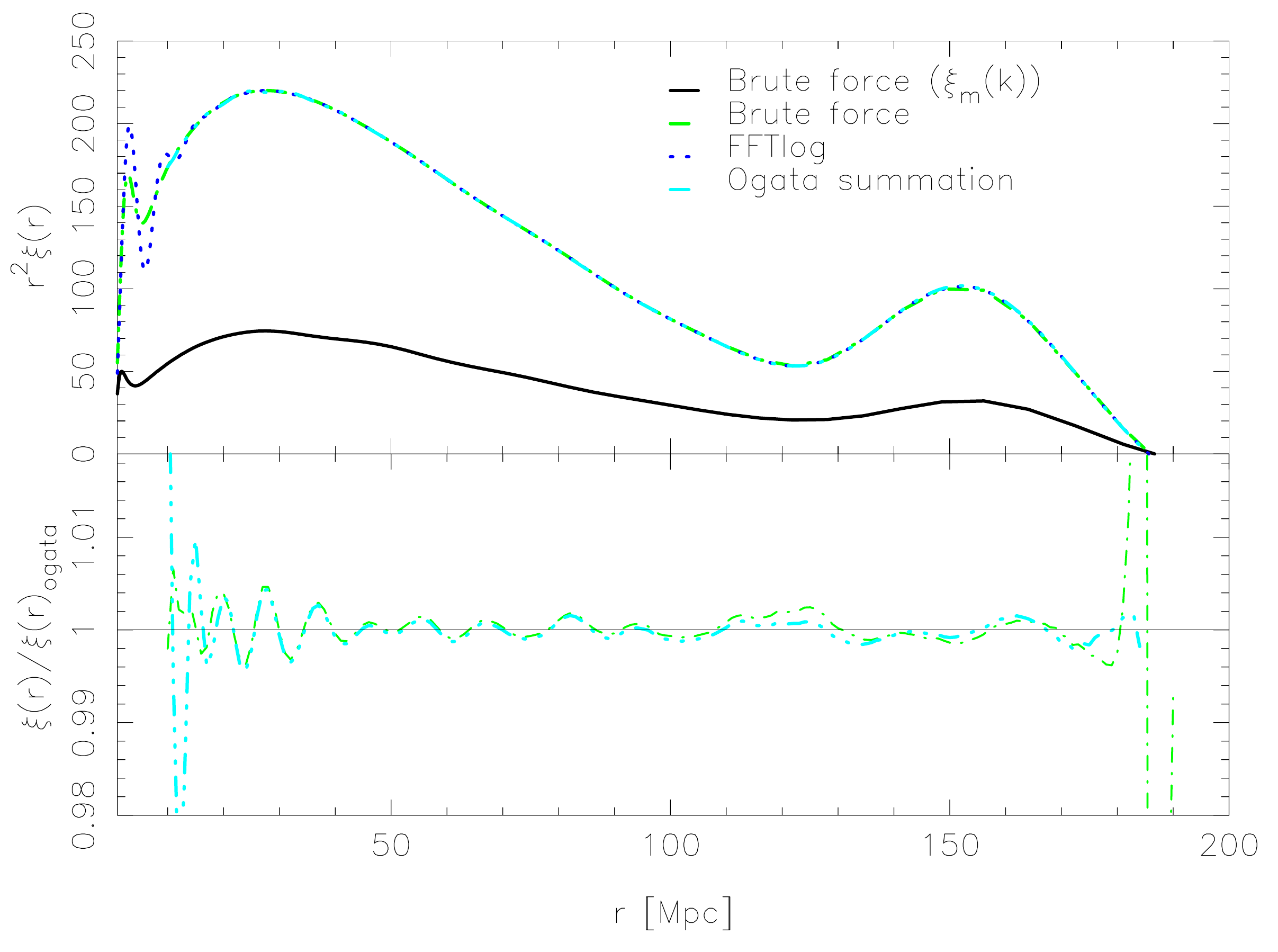} 
   \caption{Similar to Figure~\ref{fig:xi_transform_matter} but now
     testing the~\cite{ogata05} integral, FFTlog~\citep{hamilton00}
     and brute force integration, using the HOD power spectrum
     emulator as input. The power spectrum has been extrapolated into
     both the linear and nonlinear regimes by a simple power law and
     using a Pad\'{e} approximant respectively.}
   \label{fig:xi_transform_gal}
\end{figure}

As shown in~\cite{ogata05}, integrals involving Bessel functions such
as the one that appears in Equation~\ref{eqn:pktoxi} can be
approximated as:
\begin{eqnarray}
&&\int f(x) J_\nu(x)\; dx \nonumber\\
&&\approx \pi \sum_{n=1}^k w_{\nu k}f(\pi
\psi_{\nu k}(hr_{\nu k})/h) J (\pi \psi_{\nu k}(hr_{\nu k})/h)
{\psi_{\nu k}}^{\prime}(hr_{\nu k})\nonumber\\
\label{eqn:ogata}
\end{eqnarray}
where $\psi(x) = x\tanh[(\pi/2)\sinh (x)]$, $h$ is the step length of
the integration, $w_{\nu k}(x) = Y_{\nu}(x)/J_{\nu+1}(x)$ and
$r_{\nu}$ are the zeros of the Bessel function. For the correlation
function, we want to consider $\nu = 1/2$, since $j_0(x) =
\sqrt{(\pi/2x)} J_{1/2} $. Two parameters, $h$, the step size, and
$k$, the number of steps performed, control the accuracy of the
integration.  We use $h = 1/150$ and $m = 500$ and have verified that
for $f(x) = 1$, the integral is indeed equal to 1.  We found that
these parameters were a good compromise between accuracy and the time
taken to calculate the sum in Equation~\ref{eqn:ogata}.  However,
these two methods also require the power spectrum to be known over a
large range of wavelengths, much larger than the range covered by the
HOD emulator. 

To extend the power spectrum to larger length scales, we use the
smoothed matter power spectrum multiplied by the linear bias up to
$k=0.001$ Mpc$^{-1}$ and beyond that, we match a simple power law
extrapolation, $P(k) \propto k^n$, to the amplitude of the primordial
power spectrum, where $n = 0.963$.

The smaller scales are extended to $k = 10^3$ Mpc$^{-1}$ with a scheme
based on a Pad\'{e} approximant, $R(x)$, which is a series approximation
defined as $R_{[n,m]}(x) = \left(\sum^n_{i=0} a_n
  x^n\right)/\left(1+\sum^m_{i=1} b_m x^m\right)$ for constants, $a_0,
a_1, ..., a_n$, and $b_1, b_2, ... b_n$.  Fortunately, the behavior of the 
correlation function on the scales that we emulate is not affected too much 
by the power spectrum on these extremely large scales; we only require
a smooth extrapolation that approaches zero as $k$ increases. 
We chose a Pad\'{e} approximation of the form $R_{[0,m]}(k) = a_0/(1+b_1k+b_2k^2)$,
because the power spectrum appears to approach a power law, $k^{-2}$,
on small scales.  The constants $a_0, b_1, b_2$ are set by matching
the function to $P(k)$ at three points at $k \approx 0.699, 0.848,
0.995$ Mpc$^{-1}$. Because the amplitude of the power spectrum rapidly approaches
zero in the nonlinear regime, the correlation function is largely
insensitive to the exact form used to extrapolate the power spectrum
to small scales; we only require that it exists for the integral to
converge.  Nonetheless, in Figure~\ref{fig:xi_transform_matter}, we
check each of these assumptions in turn. We generate a power spectrum
up to $k = 10$ Mpc$^{-1}$ using the extended Coyote matter power
spectrum emulator~\citep{coyoteplus} and this is transformed into a
correlation function via Equation~\ref{eqn:pktoxi} using a brute force
integration.  This acts as a reference
(Figure~\ref{fig:xi_transform_matter}; black) to which we can compare
the effect of our extrapolation methods on the resultant correlation
function. In Figure~\ref{fig:xi_transform_matter}, we show two more
correlation functions whose corresponding matter power spectrum was
extrapolated to smaller scales using the Pad\'{e} approximation (red) and
additionally extended with a power law to model the linear regime
(green). We also compare the methods proposed by~\cite{ogata05} (cyan)
and~\cite{hamilton00} (dark blue). The bottom panel shows that the
relative error as a result of making these assumptions compared to the
brute force approach is less than 1\%. 
In Figure~\ref{fig:xi_transform_gal}, we demonstrate that this still
holds for the galaxy power spectra produced by our emulator.  However,
we are only able to test the various prescriptions used to evaluate
the Hankel transform in Equation~\ref{eqn:pktoxi} because the
$k-$range of the emulator is not wide enough for the brute force
method to work without some sort of extrapolation. The three methods
that we consider in Figure~\ref{fig:xi_transform_gal} all yield
results consistent to 1\%.

\section{Galaxy-Galaxy Lensing}\label{sec:gglensing}
We now demonstrate the usefulness of our emulator by applying our
predictions for the galaxy-matter cross power spectrum to estimate the 
average tangential shear produced by galaxies residing in dark matter haloes. Galaxy-galaxy
lensing involves the distortion of background galaxy images by the dark 
matter haloes of foreground galaxies. Because galaxy-galaxy lensing
is concerned with probing the halo profile of galactic sized dark matter haloes, the 
HOD model is a natural candidate for modeling the distribution of 
galaxies on small scales. Indeed, some of the strongest constraints on small scale 
structure have been derived from applying the HOD model 
to observations of galaxy-galaxy lensing.

The tangential shear from galaxy-galaxy lensing is given by~\cite{moessner98, guzik01}:
\begin{equation}
\left<\gamma_t (\theta) \right> = 6\pi\Omega_m \int d\chi \,f_\chi(\chi', \chi) \frac{n_1(\chi)}{a(\chi)}\int dk \, k \,P_{gm}(k, \chi) \, J_2(k,\theta, \chi), 
%\left<\gamma_t(\theta)\right> = 6\pi^2\left(\frac{H_0}{c}\right)^2\Omega_m\int_0^\chi_0 d\chi W_1(\chi})\frac{}
\label{eqn:gammat}
\end{equation}
where $f(\chi', \chi) = \int d\chi' n_2(\chi') \frac{\chi (\chi-\chi')}{\chi'}$ is defined as the lens efficiency, 
$\chi$ is the comoving angular distance and $a$ is the scale factor. The normalized 
distributions of foreground (lens) and background (source) galaxies are given by 
$n_1(\chi)$ and $n_2(\chi)$ respectively. 

Our emulator also calculates the excess surface density in the plane
of the lensing potential, $\Delta\Sigma$ via the following equation:
\begin{equation}
\Delta\Sigma(R) = \Sigma_{crit}\left<\gamma_t(R)\right>, 
\label{eqn:deltasigma}
\end{equation} 
where $\Sigma_{crit} = \frac{c^2}{4\pi G}\frac{D_s}{D_lD_{ls}}$ is the critical surface
density and $D$ is the angular diameter distance in proper coordinates. 

To evaluate the Hankel transform in Equation~\ref{eqn:gammat}, it is necessary to extend the 
galaxy-dark matter power spectrum to smaller scales to allow the integral to converge. Again, we adopt the Pad{\'e} 
approximation because this approach involves minimal assumptions about the high $k$ behaviour 
of the galaxy-dark matter cross power spectrum. This time, however, we add
an additional exponential damping term to prevent the bias from becoming too large
in the small scale regime. Another possible extension is to add a baryon model (such as $\propto <M>/R^2$
to represent the stellar contribution as additional lensing with a point mass) as in~\cite{velander13}
or model of sub-halo clustering with truncated NFW profiles as in~\cite{li14} instead with additional fitted parameters.

\begin{figure}[htbp]
 %  \centering
%   \hspace{-1cm}
   \includegraphics[width=1.2\linewidth] {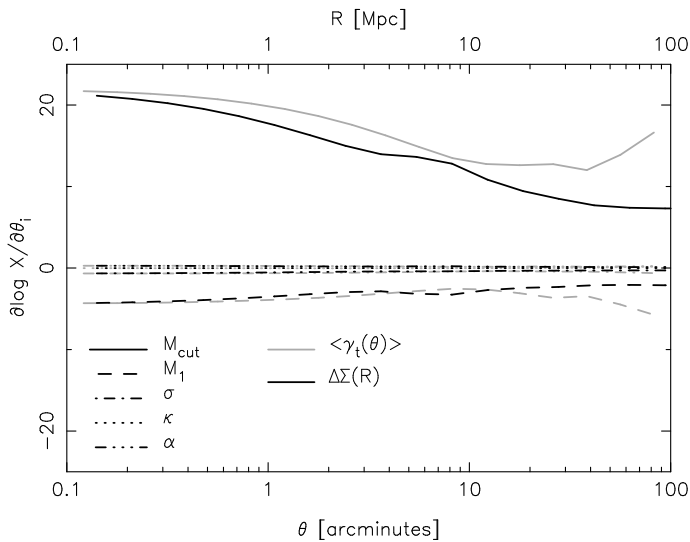}
   \caption{Derivatives of the averaged tangential shear $\gammat$ (grey)
   and excess surface density, $\DeltaSigma$ (black)
     with respect to the 5 HOD parameters, $\theta_i = \{M_{\rm cut}$ (solid), $M_1$ (dashed),
     $\sigma$ (dotted-dashed), $\kappa$ (dotted) and $\alpha$ (dotted-dotted-
     dashed) $\}$. We assume a redshift range of $0.2 < z < 0.4$ 
     for the sources and $0.5 < z < 1.3 $ for the lens catalog and a 
     redshift distribution typical of the Dark Energy Survey.}
   \label{fig:derivs_shear}
\end{figure}

In Figure~\ref{fig:derivs_shear}, we show the dependence of the tangential shear, $\gammat$,  (grey) 
excess surface density, $\DeltaSigma$, on our five HOD parameters. By far, the mass cut on centrals, 
$M_{\rm cut}$, dominates all the other HOD parameters and we can only expect to constrain two parameters,  
$M_{\rm cut}$ and $M_1$ easily. This motivates a joint analysis involving another statistic such as $w(\theta)$ 
to break the degeneracy between the HOD parameters. 
Furthermore, Figure~\ref{fig:derivs_shear} suggests that these HOD 
parameters may be more readily constrained from $\gammat$ than $\DeltaSigma$.

The use of our emulator removes the reliance on halo model based methods and provides
a significantly more accurate estimate of the non-linear clustering in the large-$k$ regime 
as demonstrated in our comparisons with halo model in Section~\ref{sec:analytic}. 
In addition, the emulator is substantially faster at evaluating the cross galaxy-matter power spectrum 
than any of the analytic methods that we have considered. 
Using the emulator instead of an analytic model reduces the run time on a typical MCMC
analysis with $\sim10^{4}$ steps for convergence from $\sim$10 hours to $\sim$ 15 minutes
on a single processor, since each evaluation of $\gammat$ with the emulator saves about five seconds.
The emulator will be applied to observations of $\gammat$ and $\DeltaSigma$
to determine the HOD of galaxies in the sample (in prep). For this purpose, we have additionally
built an emulator with an extended parameter range covering halos down to M$_{\rm cut} \sim 10^{12.5}$M$_{\odot}$
with only an additional 50 models per redshift, at the cost of downgrading the accuracy of the power 
spectra to $\sim$ 5\%\footnote{This code is available from the authors by request.}. 
Fortunately, this is not an issue for current datasets. This allows our tool to be more robust against 
changes in galaxy samples. The nested design demonstrates the flexibility of the 
Cosmic Calibration Framework as a powerful tool for providing fast, nonlinear predictions 
for the purposes of deriving cosmological constraints from observations.

%\begin{figure*}
%\centering
%\includegraphics[width = 2\columnwidth]{figures/gamma_t_sensitivity.pdf}
%\caption{Emulated $\gammat$ using the DES SVA1 distribution of lenses 
%  and sources. The five HOD parameters $M_{cut}, M_1, \sigma, \kappa$ and $\alpha$ are varied.
%  As in Figure~\ref{fig:sensitivity}, each parameter is varied one at a time while 
%  the other parameters are kept fixed at the midpoint of the parameter range.}
%  \label{fig:sensitivity_gammat}
%\end{figure*}
%
%\begin{figure*}
%\centering
%\includegraphics[width = 2\columnwidth]{figures/DeltaSigma_sensitivity.pdf}
%\caption{Emulated $\DeltaSigma$ using the DES Redmagic galaxies from SVA1. 
%  The five HOD parameters $M_{cut}, M_1, \sigma, \kappa$ and $\alpha$ are varied
%  as in Figures~\ref{fig:sensitivity} and~\ref{fig:sensitivity_DeltaSigma}. }
%  \label{fig:sensitivity_DeltaSigma}
%\end{figure*}
%

\section{Conclusions}
\label{sec:conclusion}
We present an emulator for an HOD-based galaxy-galaxy and galaxy-dark matter
cross power spectra (and correlation functions) obtained from an $N$-body simulation. 
The emulator is accurate to $\sim$1\% (auto) and $\sim$2\% (cross) over the range 
$0.01 \leq k \leq 1$~Mpc$^{-1}$ $(1 \leq r \leq 180 \; \rm{Mpc})$ from $0 \leq z \leq 1$. Using our emulator, we 
explore the parameter degeneracies of the five-parameter HOD model
of~\cite{zheng09}, finding significant degeneracies between the parameters 
on large scales. Changes in M$_{\rm cut}$ dominate both the shape and overall 
amplitude of the HOD power spectrum, while the parameter $\kappa$ has a very small effect.
We show how the emulator can be used to extract scale-dependent galaxy
bias. By comparing against the emulator results, we find that that
analytic halo model predictions for the galaxy bias, such as that of
\cite{zheng04} and~\cite{tinker05}, while reasonably accurate at large length scales
($k<0.1$~Mpc$^{-1}$), cannot be used to derive accurate constraints on
cosmology or the HOD parameters as the form of the resulting galaxy
power spectrum is significantly biased at the 20-40\% level at smaller
length scales. We also extend the emulator to provide predictions for the averaged
tangential shear, $\gammat$ and excess surface density, $\DeltaSigma$,
which are highly useful for obtaining the HOD of source galaxies from 
galaxy-galaxy lensing. We explore the parameter degeneracies in these
statistics and find that the main parameter that affects the measurement
of the shear is M$_{cut}$. 

Emulation is a powerful technique for efficiently generating accurate
models for highly nonlinear quantities in cosmology, such as the
galaxy power spectrum. The emulator only requires a small number (100)
of models to be directly computed from the halo catalog of an $N$-body
simulation, following which each prediction from the emulator takes
less than a second.  This can substantially reduce the run time needed
for an MCMC analysis where, in the current approach
e.g.~\cite{white11, parejko13}, the halo catalog and power spectrum
are recomputed at each step. To facilitate use by the community, the
emulator code has been publicly released.

Future plans to extend the emulator include the addition of
cosmological parameters to the HOD parameter space and the modeling of
redshift space distortions.

\section{Acknowledgments}

JK thanks Dave Higdon and Amol Upadhye for useful
discussions.  Partial support for JK and KH was provided by NASA. NF
and SH acknowledge partial support from the Scientific Discovery
through Advanced Computing (SciDAC) program funded by the
U.S. Department of Energy, Office of Science, jointly by Advanced
Scientific Computing Research and High Energy Physics.

This research used resources of the Argonne Leadership Computing
Facility (ALCF) under a Mira Early Science Project program. The ALCF
is supported by DOE/SC under contract DE-AC02-06CH11357. Some of the
work was conducted at the National Energy Research Scientific
Computing Center, which is supported by the Office of Science of the
U.S. Department of Energy under Contract No. DE-AC02-05CH11231.

The submitted manuscript has been created by UChicago Argonne, LLC,
Operator of Argonne National Laboratory (``Argonne''). Argonne, a
U.S. Department of Energy Office of Science laboratory, is operated
under Contract No. DE-AC02- 06CH11357. The U.S. Government retains for
itself, and others acting on its behalf, a paid-up nonexclusive,
irrevocable worldwide license in said article to reproduce, prepare
derivative works, distribute copies to the public, and perform
publicly and display publicly, by or on behalf of the Government.


\begin{thebibliography}{99}
\bibitem[{Anderson et al.}(2012)]{anderson12}
Anderson,~L., Aubourg,~E., Bailey,~S., Bizyaev,~D., Blanton,~M., 
Bolton,~A.~S., Brinkmann,~J., Brownstein,~J.~R. et al. 2012, \mnras, 427, 3435

\bibitem[{Anderson et al.}(2014)]{anderson14}
Anderson,~L., Aubourg, \'{E}, Bailey,~S., Beutler,~F., Bhardwaj,~V., Blanton,~M., 
Bolton,~A.,~S., Brinkmann,~J., Brownstein,~J.,~R. et al. 2014, \mnras, 441, 24

\bibitem[{Baugh}(2006)]{baugh06}
Baugh,~C.M. 2006, Rep. Prog. Phys., 69, 3101

\bibitem[{Baugh}(2013)]{baugh13}
Baugh,~C.M. 2013, PASA, 30, e030; arXiv:1302.2768 [astro-ph.CO]

\bibitem[{Benson et al.}(2000)]{benson00}
Benson,~A.J., Cole,~S., Frenk,~C.S., Baugh,~C.M., \& Lacey,~C.G. 2000,
MNRAS, 311, 793 

\bibitem[{Benson et al.}(2003)]{benson03}
Benson,~A.J., Bower,~R.G., Frenk,~C.S., Lacey,~C.G., Baugh,~C.M., \&
Cole,~S. 2003, \apj, 599, 38 

\bibitem[{Benson}(2010)]{benson10}
Benson,~A.J. 2010, Phys. Rep., 495, 33

\bibitem[{Berlind \& Weinberg}(2002)]{berlind02}
Berlind,~A.A. \& Weinberg,~D.H. 2002, \apj, 575, 587

\bibitem[{Bhattacharya et al. }(2013)]{bhat13}
Bhattacharya,~S., Habib,~S., Heitmann,~K., \& Vikhlinin, A, 2013, \apj, 766, 32

\bibitem[Blake et al. (2008)]{blake08}
Blake,~C., Collister,~A., \& Lahav,~O. 2008, \mnras, 385, 1257

\bibitem[{Brown et al. (2008)}]{brown08}
Brown,~M.J.I., Zheng,~Z., White,~M., Dey,~A., Jannuzi,~B.T.,
Benson,~A.J., Brand,~K., Brodwin,~M., \& Croton, ~D.J. 2008, \apj,
682, 937 

\bibitem[{Carlson et al.~(2009)}]{carlson09}
Carlson,~J.,White,~M. \& Padmanabhan,~N., Phys. Rev. D, 80, 043531

\bibitem[Cole et al. (1994)]{cole94}
Cole,~S., Aragon-Salamanca,~A., Frenk,~C.S., Navarro,~J.F., \&
Zepf,~S.E. 1994, \mnras, 271, 781

\bibitem[Cole et al. (2005)]{cole05}
Cole,~S., Percival,~W.~J., Peacock,~J.~A., Norberg,~P, B., 
Carlton,~B.,~M., Frenk,~C.,S., Baldry,~I.,~Bland-Hawthorn, J., 
et al. 2005, \mnras, 362, 505

\bibitem[{Conroy et al.}(2006)]{conroy06}
Conroy,~C., Wechsler,~R.H., \& Kravtsov,~A.V. 2006, \apj, 647, 201 

\bibitem[{Cooray \& Sheth}(2002)]{cooray02}
Cooray,~A. \& Sheth,~R. 2002, \physrep, 372, 1

\bibitem[{Crocce \& Scoccimarro}(2006)]{crocce06}
Crocce,~M. \& Scoccimarro,~R, Phys. Rev. D, 73, 063519 

\bibitem[{Davis et al.}(1985)]{davis85}
Davis,~M., Efstathiou,~G., Frenk,~C., \& White,~S.D.M. 1985, \apj,
292, 371 

\bibitem[{Dekel \& Rees}(1987)]{dekel87}
Dekel,~A. \& Rees,~M.J. 1987, Nature 326, 455

\bibitem[{Einasto et al.}(1984)]{einasto84}
Einasto,~J., Klypin,~A.A., Saar,~E., \& Shandarin,~S.F. 1984, MNRAS,
206, 529 

\bibitem[{Eisenstein et al.}(2005)]{eisenstein05}
Eisenstein,~D.J., Zehavi, I., Hogg, D. W., et al. 2005, \apj, 633, 560

\bibitem[{Gao et al.}(2005)]{gao05}
Gao,~L., Springel,~V., \& White,~S.D.M. 2005, MNRAS, 363, L66

\bibitem[{Guo et al.}(2010)]{guo10}
Guo,~Q., White,~S., Cheng,~L., \& Bolyan-Kolchin,~M. 2010, MNRAS, 404,
1111 

\bibitem[{Guzik \& Seljak}(2001)]{guzik01}
Guzik, J. \& Seljak, U, 2001, \mnras, 321, 439

\bibitem[{{Habib et al.}(2007)}]{HHHNW} 
Habib,~S., Heitmann,~K., Higdon,~D., Nakhleh,~C., \&
Williams,~B. 2007, Phys. Rev. D 76, 083503
  
\bibitem[Habib et al. (2009)]{habib09} 
Habib,~S., Pope,~A., Luki\'c,~Z., Daniel,~D., Fasel,~P., Desai,~N.,
Heitmann,~K., Hsu,~C.-H., Ankeny,~L., Mark~G., Bhattacharya,~S., \&
Ahrens,~J. 2009, J. Phys. Conf. Ser., 180, 012019

\bibitem[{{Habib et al.}(2012)}]{habib12} 
Habib,~S., Morozov,~V., Finkel,~H., Pope,~A., Heitmann,~K.,
Kumaran,~K., Peterka,~T., Insley,~J., Daniel,~D., Fasel,~P.,
Frontiere,~N., \& Luki\'c,~Z. 2012, arXiv:1211.4864 [cs.DC]

\bibitem[Hamilton (2000)]{hamilton00}
Hamilton,~A.~J.~S., 2000, \mnras 312, 257

\bibitem[{Heitmann et al.}(2006)]{HHNH} 
Heitmann,~K., Higdon,~D., Nakhleh,~C., \& Habib,~S. 2006, \apj, 646,
L1 

\bibitem[{Heitmann et al.}(2009)]{coyote2} 
Heitmann~K., Higdon~D., White~M., Habib~S., Williams,~B.J., \& Wagner,
C. 2009, \apj, 705, 156

\bibitem[{Heitmann et al.}(2010)]{coyote1}
Heitmann~K., White~M., Wagner~C., Habib~S., \& Higdon~D. 2010, \apj,
715, 104 
  
\bibitem[{Heitmann et al.}(2014)]{coyoteplus}
Heitmann,~K., Lawrence,~E., Kwan,~J., Habib,~S., \& Higdon,~D. 2014,
\apj, 780, 111

\bibitem[{Jing et al.}(1998)]{jing98}
Jing,~Y.P., Mo~H.J., \& B\"orner,~G. 1998, \apj, 494, 1

\bibitem[{Kauffmann, White, \& Guiderdoni}(1993)]{kauffmann93}
Kauffmann,~G., White,~S.D.M., \& Guiderdoni,~B. 1993, MNRAS, 264, 201  

\bibitem[{Kauffmann et al.}(1997)]{kauffmann97}
Kauffmann,~G., Nusser,~A., \& Steinmetz,~M. 1997, MNRAS, 286, 795  

\bibitem[{Kaiser}(1984)]{kaiser84}
Kaiser,~N. 1984, \apj, 284, L9

\bibitem[{Kulkarni} {et~al}(2007)]{kulkarni07}
Kulkarni,~G.V., Nichol,~R.C., Sheth,~R.K., Seo, ~H.-J.,
Eisenstein,~D.J., \& Gray,~A. 2007, \mnras, 378, 1196

\bibitem[{Kwan et al.}(2013)]{kwan13}
Kwan,~J., Bhattacharya,~S., Heitmann,~K., \& Habib,~S. 2013, \apj,
768, 123 

\bibitem[{{Lawrence et al.}(2010)}]{coyote3} Lawrence,~E.,
Heitmann,~K., White~M., Higdon~D., Wagner~C., Habib~S., \& Williams,
B. 2010, \apj, 713, 1322 

\bibitem[{Li et al.}(2014)]{li14}
Li., R, Shan, H., Mo, H., et al., 2014, \mnras,d 438, 2864

\bibitem[{Lukic et al}(2007)]{lukic07}
Lukic,~Z., Heitmann,~K., Habib,~S., Bashinsky,~S., \& 
Ricker,~P.,~M., 2007, \apj, 671, 1160

\bibitem[McDonald (2006a)]{mcdonald06}
McDonald,~P., 2006, \prd, 74, 103512

\bibitem[McDonald (2006b)]{mcdonald06err}
McDonald,~P., 2006, \prd, 74, 129901

\bibitem[{Moessner \& Jain} (1998)]{moessner98}
Moessner, R. \& Jain, B, 1998, \mnras, 294, L18

\bibitem[Moster et al. (2010)]{moster10}
Moster,~B.P., Somerville,~R.S., Maulbetsch,~C., van den Bosch,~F.C.,
Maccio,~A.V., Naab,~T., \& Oser,~L. 2010, \apj, 710, 903

\bibitem[Neistein and Khochfar (2012)]{neistein12}
Neistein,~E. \& Khochfar, S., 2012, arxiv:1209.0463

\bibitem[{Ogata} (2005)]{ogata05}
Ogata,~H., 2005, Publ. RIMS, Kyoto University, 41, 949

\bibitem[{Padmanabhan} {et~al.}(2009)]{padmanabhan09}
Padmanabhan,~N., White,~M., Norberg,~P., \& Porciani,~C., 2009,
\mnras, 397, 1862

\bibitem[Parejko et al.~(2013)]{parejko13}
Parejko,~J.K., Sunayama,~T., Padmanabhan,~N., Wake,~D.,~A., 
Berlind,~A.,~A., Bizyaev,~D., Blanton,~M., Bolton,~A.,~S. et al. 2013, \mnras, 428, 98  

\bibitem[{Parkinson et al.}(2012)]{parkinson12}
Parkinson,~D., Riemer-Sorensen,~S., Blake,~C., Poole,~G.,~B., Davis,~T.,~M., 
Brough,~S., Colless,~M., Contreras,~C., et al. 2012, Phys, Rev. D, 86, 103518

\bibitem[Peacock \& Smith (2000)]{peacock00}
Peacock,~J.~A., \& Smith.,~R.E. 2000, \mnras, 318, 1144

\bibitem[{Phleps} {et~al.}(2006)]{phleps06}
Phleps, S., Peacock,~J.~A., Meisenheimer, K., \& Wolf, C., 2006, \aap,
457, 145 

\bibitem[{Pope et al.}(2004)]{pope04}
Pope,~A.~C., Matsubara,~T., Szalay,~A.,~S., Blanton,~M.,~R., 
Eisenstein,~D.,~J., Gray,~J., Jain,~B., Bahcall,~N.,~A. et al. 2004, \apj, 607, 655 

\bibitem[Pope et al.~(2010)]{pope10}
Pope,~A.~C, Habib,~S, Luki\'c,~Z., Daniel,~D., Fasel,~P., Desai,~N., \&
Heitmann,~K. 2010,  Comp. Sci. \& Eng. 12, 17  

\bibitem[{Reid et al.}(2012)]{reid12} 
Reid,~B.A., Samushia,~L., White,~M., Percival,~W.,~J., Manera,~M., 
Padmanabhan,~N., Ross,~A.,~J., S\'{a}nchez,~A.,~G. et al. 2012, \mnras, 426, 2719 

\bibitem[{Reid et al.}(2014)]{reid14} 
Reid,~B.A., Seo~,H.-J., Leauthaud,~A., Tinker,~J.L. \& White,~M. 2014,
\mnras, 444, 476

\bibitem[Seljak (2000)]{seljak00}
Seljak, U. 2000, \mnras, 318, 203 

\bibitem[Somerville \& Primack (1999)]{somerville99}
Somerville, R.S., \& Primack,~J.R. 1999, \mnras, 310, 1087 

\bibitem[{Swanson~et~al.}(2010)]{swanson10}	
Swanson,~M.~E.~C., Percival, ~W.~J. \& Lahav,~O, 2010, \mnras, 409, 1100

\bibitem[Szapudi et~al. (2005)]{szapudi05}
Szapudi, I., Pan, J.,  Prunet, S., \& Budav\'ari, T., 2005, \apj, 631, 1

\bibitem[{Tinker~et~al.}(2005)]{tinker05}
Tinker, J., Weinberg, D., Zheng, Z., \& Zehavi, I., 2005, \apj, 631,  41

\bibitem[{Tinker~et~al.}(2008)]{tinker08}
Tinker, J., Kravtsov, A. V., Klypin, A., Abazajian, K., Warren, M., 
Yepes, G.,  Gottl\"{o}ber, S., \& Holz, D. E., 2008, \apj, 688,  709

\bibitem[{Tinker~et~al.}(2010)]{tinker10}
Tinker, J., Robertson,~B.,~E. et al., Kravtsov, A. V., Klypin, A., 
Warren,~M.,~S., Yepes,~G., \& Gottl\"{o}ber, S., 2010, \apj, 724,  878

\bibitem[Tegmark~(1997)]{tegmark97}   
Tegmark,~M. 1997, \prl 79, 3806

\bibitem[{Tegmark et al.}(2004)]{tegmark04}
Tegmark,~M., Blanton,~M.,~R., Strauss,~M.,~A., Hoyle,~F., Schlegel,~D., 
Scoccimarro,~R., Vogeley,~M.,~S., Weinberg,~D.,~H. et al. 2004, \apj, 606, 702

\bibitem[{Tegmark et al.}(2006)]{tegmark06}
Tegmark,~M., Eisenstein,~D.,~J., Strauss,~M.,~A., Weinberg,~D.,~H., 
Blanton,~M.,~R., Frieman,~J.,~A., Fukugita,~M., Gunn,~J.,~E., et al. 2006, Phys. Rev. D, 74, 123507

\bibitem[{Vale \& Ostriker}(2004)]{vale04}
Vale,~A. \& Ostriker,~J.P. 2004, MNRAS, 353, 189

\bibitem[{Velander} {et~al.}(2013)]{velander13}
Velander,~M., van Uitert, E, Hoekstra, H., et al., 2013, \mnras, 437, 2111
 
\bibitem[{Wake} {et~al.}(2008)]{wake08}
Wake,~D.A., Sheth,~R.,~K., Nichol,~R.,~C., Baugh,~C.,~M., Bland-Hawthorn,~J., 
Colless,~M., Couch,~W.,~J., Croom,~S.~M. et al. 2008, \mnras, 387, 1045  

\bibitem[Wetzel \& White (2010)]{wetzel10}
Wetzel,~A.R., \& White,~M. 2010, \mnras, 403, 1072  
 
\bibitem[White et al.~(2011)]{white11}
White,~M., Blanton,~M.,  Bolton,~A., Schlegel,~D., Tinker,~J., 
Berlind,~A., da Costa,~L., Kazin, E. et al. 2011, \apj, 728, 126

\bibitem[White \& Frenk~(1991)]{white91}
White,~S.D.M., \& Frenk,~C.S. 1991, \apj, 379, 52

\bibitem[Zehavi et al. (2004)]{zehavi04}
Zehavi,~I., Weinberg,~D.,~H., Zheng,~Z., Berlind,~A.,~A., Frieman,~J.,~A., 
Scoccimarro,~R., Sheth,~R.~K. Blanton,~M.,~R., et al. 2004, \apj, 608, 16

\bibitem[Zehavi et~al. (2005)]{zehavi05}
Zehavi,~I, Zheng,~Z.,  Weinberg,~D.,~H., Frieman,~J.,~A., Berlind,~A., A., 
Blanton,~M.,~R., Scoccimarro,~R., Sheth,~R., K., et al. 2005, \apj, 630, 1

\bibitem[{Zheng}(2004)]{zheng04}
Zheng,~Z, 2004, \apj, 610, 61

\bibitem[{Zheng et al.}(2005)]{zheng05}
Zheng,~Z., Berlind,~A.,~A., Weinberg,~D.,~H., Benson,~A.,~J., Baugh,~C.,~M., 
Cole,~S., Dav\'{e},~R., Frenk,~C.,~S., et al. 2005, \apj, 633, 791

\bibitem[{Zheng et al.}(2009)]{zheng09}
Zheng,~Z, Zehavi,~I., Eisenstein,~D.J.; Weinberg,~D.H., \&
Jing,~Y.P. 2009, \apj, 707, 554 


\end{thebibliography}
\end{document}